\documentclass[aps,prd,nofootinbib,twocolumn,superscriptaddress,preprintnumbers,balancelastpage,longbibliography]{revtex4-1}

\usepackage{bm}
\usepackage{bbm}
\usepackage{appendix}
\usepackage[utf8]{inputenc}
\usepackage{amsmath,amssymb,mathtools,bm}
\usepackage{graphicx, color, hepunits}
\usepackage[dvipsnames]{xcolor}
\usepackage{float}
\usepackage{multirow}
\usepackage{tikz-feynman} 
 \usepackage{hyperref}
\hypersetup{
    colorlinks=true,       
    linkcolor=blue,        
    citecolor=blue,        
    filecolor=magenta,     
    urlcolor=blue        
}
\usepackage[utf8]{inputenc}
\usepackage[english]{babel}
\let\hvec\vec
\let\vec\mathbf
\usepackage{tensor}
\usepackage{orcidlink}
\usepackage{tikz}
\usepackage{graphicx}
\usepackage{subcaption}
\usetikzlibrary{calc,arrows.meta,fit,backgrounds}
\renewcommand{\eqref}[1]{Eq.~(\ref{#1})}

\newcommand{\ket}[1]{\left|#1\right\rangle}    
\newcommand{\bra}[1]{\left\langle#1\right|}    
\newcommand{\tj}[6]{ \begin{pmatrix}
  #1 & #2 & #3 \\
  #4 & #5 & #6 
 \end{pmatrix}} 

\begin{document}

\title{A Numerical Method for the Efficient Calculation of Scattering Form Factors}

\author{Benjamin Lillard~\orcidlink{0000-0001-8496-4808}}
\affiliation{Institute for Gravitation and the Cosmos, The Pennsylvania State University, University Park, PA 16802, USA}
\affiliation{{Department of Physics, University of Oregon, Eugene, OR, 97403, USA}}

\author{Jack D. Shergold~\orcidlink{0000-0001-5685-9007}}

\affiliation{{Department of Mathematical Sciences, University of Liverpool, Liverpool, England}}

\author{Carlos Blanco~\orcidlink{0000-0001-8971-834X}}
\affiliation{Institute for Gravitation and the Cosmos, The Pennsylvania State University, University Park, PA 16802, USA}
\affiliation{Department of Physics, Princeton University, Princeton, NJ 08544, USA}
\affiliation{Stockholm University and The Oskar Klein Centre for Cosmoparticle Physics, Alba Nova, 10691 Stockholm, Sweden}

\date{\today}

\begin{abstract}
Scintillating molecular crystals have emerged as prime candidates for directional dark matter detector targets. This anisotropy makes them exquisitely sensitive due to the daily modulation induced by the directional dark matter wind. However, predicting the interaction rate for arbitrary molecules requires accurate modeling of the many-body ground as well as excited states, a task that has been historically computationally expensive. Here, we present a theory and computational framework for efficiently computing dark matter scattering form factors for molecules. We introduce \texttt{SCarFFF}, a GPU-accelerated code to compute the fully three-dimensional anisotropic molecular form factor for arbitrary molecules. We use a full time-dependent density functional theory framework to compute the lowest-lying singlet excited states.
Once the many-body electronic structure is computed, the form factors are computed in a small fraction of the time from the transition density matrix. We show that \texttt{ScarFFF} can compute the first 12 form factors for a molecule of 10 heavy atoms in approximately 5 seconds, opening the door to accurate, high-throughput material screening for optimal directional dark matter detector targets. Our code can perform the calculation in three independent ways, two semi-analytical and one fully numeric, providing optimized methods for every precision goal.  
\end{abstract}
\maketitle


\section{Introduction}
\label{sec:Introduction}

For thousands of years, scientists have studied the properties of matter by transferring momentum and energy into various systems.  Our knowledge of the universe has been greatly advanced through a cycle of increasingly precise predictions and measurements, addressing the age-old question---\emph{what will happen if I hit it?}---across a vast range of energies and length scales.
Currently, all of the types of matter we can manipulate or detect are described precisely by the Standard Model (SM) of particle physics. The primary exception is dark matter, which interacts with SM matter gravitationally~\cite{Zwicky:1933gu,Rubin:1970zza}, but has not yet been conclusively detected through other means. The particle description of dark matter (DM) remains one of the outstanding mysteries in modern physics.

In order to discover new particles, we need a precise understanding of our tools: namely, the SM systems that constitute our detectors. In the case of DM direct detection experiments, we need to know how a  material responds when a DM particle imparts some momentum $(\vec q)$ and energy $(E)$ onto SM particle. Except for the simplest single-particle systems, predicting this SM material property is computationally expensive, especially when the material is anisotropic, i.e.~not spherically symmetric. The difficulty of predicting the material response functions has limited the development of new detector designs, particularly for DM masses below 1~GeV. 

In this work we use a combination of analytic and numerical methods to overcome this limitation for DM--electron scattering. Our Julia package, \texttt{SCarFFF}\footnote{\texttt{SCarFFF}: \url{https://github.com/jdshergold/SCarFFF}} (Spherical, Cartesian, and Fourier Form Factors), can calculate new molecular form factors from first principles in $\lesssim 4\,\mathrm{s}$ per transition, for molecules around $10$ heavy atoms.
Using \texttt{SCarFFF} to find the form factor and Vector Spaces for Dark Matter (\texttt{VSDM}), a recently published package for efficiently calculating the DM scattering rates~\cite{Lillard:2023qlx,Lillard:2023cyy,Lillard:2025aim}, to calculate the DM scattering rate, a researcher can complete a daily modulation analysis including $\mathcal O(10)$ excited states in about one minute per new molecule.

\onecolumngrid

\begin{figure}[h]
\centering
\includegraphics[width=0.98\textwidth]{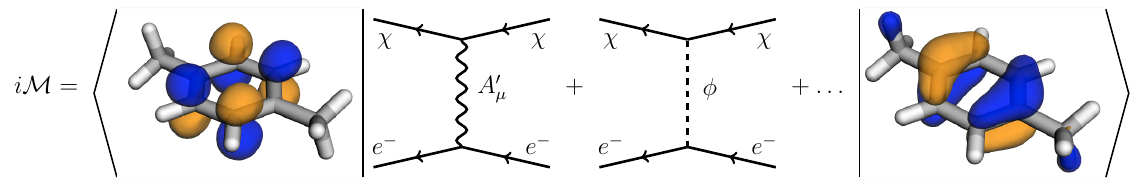}
\onecolumngrid
\caption{
    A weakly interacting particle $\chi$ scatters with an electron in a multiparticle ground state, transferring to it some momentum and energy, and leaving it in a higher energy eigenstate.
} \label{fig:amplitude}
\end{figure}

\twocolumngrid
Our simplest method, the discrete Fourier transform of Section~\ref{sec:fft}, can be used in any system where the initial and final state electronic wavefunctions can be tabulated in position space. It can be the fastest way to calculate the form factor when imprecise results are tolerated. Our fastest analytic method, unveiled in Section~\ref{sec:analytic}, is specific to physical chemistry calculations performed using a basis of Gaussian-type orbitals. Unlike the FFT, it can be performed without any loss of precision.

Currently, \texttt{SCarFFF} calculates the scalar molecular form factor, a.k.a.~the dynamic structure factor or electron loss function, which is sufficient for spin-independent DM--electron scattering~\cite{Trickle:2019nya,Hochberg:2021pkt,Knapen:2021run,Lasenby:2021wsc,Boyd:2022tcn}. A future version of \texttt{SCarFFF} will also calculate the spin response functions necessary for spin-dependent DM interactions~\cite{Catena:2024rym,Berlin:2025uka,Hochberg:2025rjs,Giffin:2025hdx}.

\section{Electron Scattering Calculation}

\subsection{Spin-Independent Interaction Rate}
The rate of DM--$e^-$ scattering events in a material is given by an integral over the DM velocity $\vec v$ and the momentum transfer $\vec q$~\cite{Essig:2015cda,Blanco:2021hlm}:
\begin{align}
R_{g \rightarrow s}  &= \frac{N_\text{cell} \rho_\chi /m_\chi}{128 \pi^2 m_\chi^2 m_e^2} \int\! \frac{d^3 q}{q} \, \eta(\vec q, E)
\left| \mathcal M_{g \rightarrow s} (\vec q) \right|^2 ,
\label{eq:rate}
\\
\eta(\vec q, E) &\equiv 2q \int\! d^3v \, g_\chi(\vec v) \, \delta\!\left( E+ \frac{q^2}{2 m_\chi} - \vec q \cdot \vec v \right) 
\end{align}
where $m_\chi$ and $\rho_\chi$ are the DM mass and local density, $m_e$ is the electron mass, $N_\text{cell}$ is the number of target particles or unit cells in the detector. Following Ref.~\cite{Lillard:2025aim}, 
$\eta(\vec q, E)$ is the integrated DM velocity distribution, defined in terms of the lab-frame velocity distribution $g_\chi(\vec v)$, which satisfies $\int d^3v \,g_\chi(\vec v) \equiv 1$. 

Our label $g\rightarrow s$ refers to a transition from the (multiparticle) electronic ground state $\ket{\Psi_g}$ to a specific excited state $\ket{\Psi_s}$.
The scattering amplitude can be written as
\begin{align}
\mathcal M_{g \rightarrow s} ( \vec q) &= \langle \Psi_s | \delta H | \Psi_g \rangle,
\label{def:M12}
\end{align}
where  $\delta H$ is an operator encoding the DM--electron interaction. 
The details of the electronic bound state can be factorized from the DM-$e^-$ coupling into a free particle amplitude $\mathcal M_\text{free}$ and a material-specific form factor $f_s(\vec q)$, 
\begin{align}
\mathcal M_{1 \rightarrow 2} ( \vec q) &= \mathcal M_\text{free}(q) \, f_s(\vec q), 
\end{align}
where in the notation of Refs.~\cite{Trickle:2019nya,Hochberg:2021pkt,Knapen:2021run,Lasenby:2021wsc,Boyd:2022tcn} we define
\begin{align}
f_s(\vec q) &\equiv \langle \Psi_s | \tilde n_e(- \vec q) | \Psi_g \rangle
\label{def:fS}
\end{align}
where $\tilde n_e(- \vec q)$ is the Fourier transform of the electronic density operator.
It is related to the dynamic structure factor $S(\vec q, \omega)$ of Ref.~\cite{Trickle:2019nya} via~\cite{Lillard:2025aim}
\begin{align}
S(\vec q, \omega) &= \frac{2\pi}{V_\text{cell} } f_s^2(\vec q) \, \delta(\omega - E_s),
\end{align}
where $S$ is normalized by $V_\text{cell}$, the volume of the unit cell. 
For simple systems that are well described by a single particle, the form factor takes the form 
\begin{align} \label{eq:fSsingle}
f_s(\vec q) &= \int\! \frac{ d^3 k }{(2 \pi)^3} \tilde \psi_s^\star(\vec k + \vec q) \, \tilde \psi_g(\vec k) ,
\end{align}
where $\tilde\psi_i$ are the momentum space wavefunctions of the initial and final states. As we show in Section~\ref{sec:structure}, the form factor for more complicated multiparticle systems can often be put into this form.

It is typical to write $\mathcal M_\text{free}(q)$ in terms of a $q$ dependent form factor $F_\text{DM}^2$, 
\begin{align}
F_\text{DM}^2(q) \equiv \frac{ \left| \mathcal M_\mathrm{free}(q) \right|^2}{  \left| \mathcal M_\mathrm{free}(\alpha m_e) \right|^2} , 
\end{align}
and a constant cross section $\bar\sigma_e$ that parameterizes the strength of the DM--SM coupling:
\begin{align}
\bar\sigma_e &\equiv \frac{\mu_{\chi e}^2  \left| \mathcal M_\mathrm{free}(\alpha m_e) \right|^2 }{16 \pi m_e^2 m_\chi^2}.
\end{align}
With these replacements, the rate integral takes the compact form:
\begin{align} \label{eq:rateR}
R_{1 \rightarrow 2} &= \frac{N_\text{cell} \rho_\chi \bar\sigma_e}{8 \pi m_\chi \mu_{\chi e}^2 } \int\! \frac{d^3 q}{q} \eta(\vec q, E) \, F_\text{DM}^2(q) \, |f_s(\vec q) |^2 .
\end{align}

Figure~\ref{fig:amplitude} summarizes the physical system. The interaction between dark matter (or a Standard Model particle) and an electron in a momentum eigenstate can be calculated using Feynman diagrams for free particles, 
yielding the $\mathcal M_\text{free}$ contribution to the amplitude. 
The probability of finding an electron of momentum $\vec k$ in the initial state, and an electron of momentum $\vec k' = \vec k + \vec q$ in the final state, is calculated using methods from physical chemistry.

\subsection{Electronic Structure Calculation: Single Particle Orbitals} \label{sec:structure}

To determine $f_s(\vec q)$, we must begin with an accurate description of the initial and final multi-electron states. 
We work in a simple LCAO model, where single-particle molecular orbitals are linear combinations of atomic orbitals $\phi_{\alpha I}(\vec{r}) = \sum_{\alpha} c_\alpha \phi_\alpha(\vec{r}-\vec{r}_{I})$ centered on the nuclear equilibrium locations $\vec{r}_{I}$.
In our discussion of the multielectron wavefunctions, the Hartree-Fock ground state is a useful starting point. It is a spin-singlet Slater determinant of $n$ electrons in the first $n' = n/2$ single-particle molecular orbitals (MOs), ordered by energy:
\begin{align}
\ket{\psi_0} &=|\varphi_1\bar\varphi_1\ldots \varphi_{n^\prime}\bar{\varphi}_{n^\prime}|,
\end{align}
where $|\varphi_i \ldots \bar\varphi_j |$ indicates the antisymmetric Slater determinant, with $\varphi$ and $\bar\varphi$ denoting the two electronic spin states.
The first $i \leq n'$ MOs are usually referred to as ``occupied'', and the remaining $a \geq n'$ orbitals ``virtual'' or ``unoccupied.'' 
In this notation a singly excited state $\ket{\psi_i^a}$ replaces an occupied MO ($\varphi_i$) with a virtual one ($\varphi_a$). 
Since our primary focus is on prompt fluorescence in organic scintillators, 
we restrict our attention to the singlet configurations: that is,
\begin{align}
\ket{\psi_{i}^{a}}&= \frac{1}{\sqrt{2}}|\varphi_1\bar\varphi_1\ldots \varphi_i\bar\varphi_a \ldots \varphi_{n^\prime}\bar\varphi_{n^\prime}|
\nonumber\\&~~ - \frac{1}{\sqrt{2}} |\varphi_1\bar\varphi_1 \ldots \varphi_a\bar\varphi_i \ldots \varphi_{n^\prime}\bar\varphi_{n^\prime}| ,
\\
\ket{\Psi_s} &\simeq \sum_{i,a} \ket{ \psi_{i}^a } \langle \psi_i^a | \Psi_s \rangle.
\end{align}
Restricting our calculation to singlet states, we can relate $\ket{ \psi_i^a}$ to the $\ket{\psi_0}$ state via creation and annihilation operators $c_p^\dagger$ and $c_p$:
\begin{align}
\ket{\psi_i^a} &= c_a^\dagger c_i \ket{\psi_0} .
\end{align}
Triplet excited states are associated with the delayed fluorescence component of the scintillation light, i.e.~phosphorescence, which is generally much less efficient than the prompt fluorescence signal.
For this reason we will postpone a detailed treatment of triplet states to future work. 

Because the Hartree-Fock $\ket{\psi_0}$ state does not include effects from electron correlations, it is not a precise description of the actual ground state $\ket{\Psi_g}$. Instead, we allow $\ket{\Psi_g}$ to receive contributions from doubly-excited configurations of the form $\ket{\psi_{ab}^{ij} }$, 
where two of the occupied orbitals have been ``promoted'' to $a,b > n'$. 
In this basis, 
\begin{align}
\ket{\Psi_g } \simeq (1 - \epsilon^2) \ket{\psi_0} + \sum_{ij,ab} \ket{\psi_{ij}^{ab} } \langle \psi_{ij} | \Psi_g \rangle , 
\end{align}
where $\epsilon^2$ is a normalization factor.

We adopt a molecular orbital model where transitions between the ground state and excited states are found to be roots of the Casida Equation~\cite{casida1995time}. Specifically, \texttt{SCarFFF} makes use of the linear-response time-dependent density functional theory (TD-DFT) framework. Therefore, both the positive ($X_{ia}$) and negative ($Y_{ia}$) eigenvalue solutions of the Casida equation are included,
combining to form the quasiboson excitation operator:
\begin{align}
    Q_s = \sum_{\sigma,\tau}\sum_{ia} X_{ia;\sigma\tau}^{(s)} c^\dagger_{a,\tau} c_{i,\sigma} - Y_{ia;\sigma\tau}^{(s)} c^\dagger_{i,\sigma} c_{a,\tau},
\end{align}
which acts on the ground state to produce an excited state, $Q_s \ket{\Psi_g} = \ket{\Psi_s}$, where $\sigma,\tau \in \{\alpha,\beta\}$ label the two spin channels. We specialize to singlet--singlet transitions using a restricted Kohn--Sham (RKS) reference, for which the amplitudes are spin-diagonal and identical in the two spin channels: $X_{ia;\sigma\tau} = \delta_{\sigma\tau} X_{ia}$, and analogous for $Y$. The normalization includes both spin channels, such that $2\sum_{ia} (X_{ia}^2 - Y_{ia}^2) = 1$. Heuristically, the $X_{ia}$ matrix characterizes the single excitation character of the transition, while $Y_{ia}$ characterizes the correlation effects and the double excitation character of the ground state. Taking $Y\rightarrow 0$ is equivalent to the Tamm-Dancoff approximation (TDA) in linear-response theory. 
The $N_o \times N_v$ matrices $X$ and $Y$ in the MO basis can be combined into a single $(N_o + N_v)^2$ square matrix in the AO basis, $T^{(s)}_{pq}$,
\begin{align}
    T_{pq}^{(s)}&=  \sum_{ia}C_{pi}^{\mathrm{occ}}\left[X^{(s)}_{ia} + Y^{(s)}_{ia}\right]C_{qa}^{\mathrm{vir}},
\end{align}
where $C_{pi}^{\mathrm{occ}}$ and $C_{qa}^\mathrm{vir}$ are the MO coefficient matrices that rotate from the AO basis to the diagonal MO basis.

Compared to the single-particle $f_s$ of \eqref{eq:fSsingle}, it is no longer straightforward to identify a single pair of initial and final states.
For visualizing the transition in Figure~\ref{fig:amplitude}, we used the singular value decomposition of $T^{(s)}_{pq}$ to identify the natural transition orbitals of Ref.~\cite{nto_martin}.
\texttt{SCarFFF} works with $T^{(s)}_{pq}$ directly in the atomic orbital basis.

Following our previous work, the relevant single-particle observables of the many-body wavefunction can be expressed as a sum of single-particle matrix elements. The scattering form factor, characterizing the $s$-th singlet transition, is given by:
\begin{align}
    f_{s}(\textbf{q}) = 2\, T^{(s)}_{ia} \bra{\varphi_a (\textbf{r})}e^{i\textbf{q}\cdot\textbf{r}}\ket{\varphi_i(\textbf{r})},
\end{align}
where the factor of 2 in the form factor comes from summing over the identical contributions from each spin channel. Our molecular orbitals in \texttt{SCarFFF} are expanded in a basis of Gaussian-type orbitals, such as 6-31G*, cc-pVDZ or cc-pVTZ. More generally, \texttt{SCarFFF} currently supports several basis sets, constructed from up to $i$ orbitals. 

We have adopted specific exchange functionals and basis sets as a benchmark for the results of this study. However, the functionality of \texttt{SCarFFF} depends only on the availability of a transition density matrix and a selection of any Gaussian basis set. We suggest using the 6-31g$^*$ basis set to obtain qualitatively correct results, and double or triple zeta basis sets for quantitatively robust results, e.g.~cc-pVDZ or cc-pVTZ. See Appendix~\ref{appx:basis} for a detailed discussion of our basis-set conventions.

\section{FFT Method} \label{sec:fft}

The molecular form factor $f_s(\vec q)$ can be calculated using either position space or momentum space wavefunctions. 
Following Section~\ref{sec:structure},
\begin{align}
f_s(\vec q) &=\int\! d^3 r \, e^{i \vec q \cdot \vec r} \Phi_{g \rightarrow s}(\vec r),
\label{eq:fSreal}
\\
\Phi_{g \rightarrow s}(\vec r) &=  2 \sum_{pq}  T_{pq}^{(s)} \phi_q^\star(\vec r) \, \phi_p(\vec r)
\label{eq:PhiTDM}
\end{align}
where $T_{pq}^{(s)}$ is the transition density matrix for the $g \rightarrow s$ excitation in the basis of atomic orbitals ($\phi_{p,q}$).
With this notation, $f_s(- \vec q)$ is simply the Fourier transform of $ \phi T \phi(\vec r)$. 

It would be prohibitively time-consuming to evaluate $f_s(\vec q)$ by integrating \eqref{eq:fSreal} at each point $\vec q$ in momentum space. However, there are highly efficient numerical routines for approximating the Fourier transform of a tabulated $d$ dimensional function, namely the discrete Fast Fourier Transform (FFT)~\cite{FrigoJohnson2005FFTW3}. 
We find that the FFT is a viable method for extracting $f_s(\vec q)$ from tabulated transition density functions, especially when high precision is not required.

\subsection{The (Fast) Fourier Transform}
We define the forward and backward Fourier transforms as, respectively: 
\begin{align}
\mathcal F[f](\vec q) &= \int\! d^3 x\, e^{-i \vec q \cdot \vec x} f(\vec x) ,
\\
\mathcal F^{-1}[\tilde f](\vec x) &= \int \! \frac{d^3 k}{(2\pi)^3} e^{+i \vec k \cdot \vec x} \tilde f(\vec k) .
\end{align}
With this notation, $f_s$ is simply 
\begin{align}
f_s^\star(\vec q) &= \mathcal F[\Phi^\star_{g \rightarrow s} ](\vec q), 
\end{align}
or equivalently $f_s( \vec q) = \mathcal F[\Phi_{g \rightarrow s} ](- \vec q)$.

Compare this to the discrete Fourier transform. 
In three dimensions, we define the forward FFT of a 3d array $f_{a b c} = f(x_a, y_b, z_c)$ as:
\begin{align}
F_{\alpha \beta \gamma} &\equiv \sum_{a=0}^{N_x - 1} \sum_{b = 0}^{N_y - 1} \sum_{c = 0}^{N_z - 1} f_{abc} \exp\left( - 2\pi i \left( \tfrac{a \alpha }{N_x} + \tfrac{ b \beta }{N_y} + \tfrac{c \gamma}{N_z} \right) \right) ,
\label{eq:Falbega}
\end{align}
for integers $\alpha = 0, 1, \ldots, N_x - 1$, $\beta = 0 , 1, \ldots N_y- 1$, etc.

Let us tabulate $f(\vec x)$ on a regularly spaced rectangular grid, within a volume $V = L_x L_y L_z$ bounded by
\begin{align}
x \in [x_\text{min} , x_\text{max} ],
\,
y \in [y_\text{min} , y_\text{max} ], 
\,
z \in [z_\text{min} , z_\text{max} ], 
\end{align}
where we define $L_i \equiv x_\text{max}^{(i)} - x_\text{min}^{(i)}$ as the length of each dimension of the box. 
Defining 
\begin{align}
u_i &\equiv \frac{x_i - x_\text{min}^{(i)} }{L_i} , 
&
x_i &= L_i u_i + x_\text{min}^{(i)}, 
\end{align}
the Fourier transform of $f(\vec x)$ is:
\begin{align}
F(\vec k) &= \int_{V} \! d^3x \, e^{-i \vec k \cdot \vec x} f(\vec x) \notag
\\
&\simeq \sum_{a b c}  V_\text{pix} \exp\left( -i \vec k \cdot \left[ L_i u_i + x_\text{min}^{(i)} \right] \right) f_{a b c} , 
\label{eq:continuFFT}
\end{align}
where $f_{a b c}$ is the value of $f(\vec x)$ at the grid position $\vec x = \vec x_{a b c}$, defined via
\begin{align}
\vec u_{a b c} &= \frac{a}{N_x} \hat x + \frac{b}{N_y} \hat y + \frac{c}{N_z} \hat z  ,
\end{align}
with $V_\text{pix}$ the volume of a single pixel in the grid,
\begin{align}
V_\text{pix} &= \frac{L_x L_y L_z}{N_x N_y N_z}.
\end{align}
Pulling out the overall phase from the translation in $\vec x \rightarrow \vec u$, 
\begin{widetext}
\begin{align}
F(\vec k) &\simeq V_\text{pix} e^{-i \vec k \cdot \vec x_\text{min} } \sum_{a b c} f_{a b c} \exp\left( - 2\pi i \left[ \frac{k_x L_x}{2 \pi} \frac{a }{N_x} + \frac{ k_y L_y}{2 \pi} \frac{b}{N_y} + \frac{k_z  L_z}{2\pi } \frac{c}{N_z} \right] \right) .
\end{align}
\end{widetext}
Standard FFT routines, given the array $f_{abc}$ as an input, will return $F_{\alpha \beta \gamma}$. From the $2\pi$ periodicity of the complex phase in \eqref{eq:Falbega}, note that 
$(N_x - \alpha)/N_x$ describes the same Fourier mode as $- \alpha / N_x$, 
and so the FFT output needs to be rearranged by cycling the arrays until the most negative frequencies $(\alpha \geq N_x / 2)$ appear first in the list. 
Once this is done, all we need to do is multiply $F_{\alpha \beta \gamma}$ by the pixel volume, $V_\text{pix}$, and the phase associated with translation from the origin to the corner of the grid:
\begin{align}
F(k_x, k_y, k_z) &= F\left( \frac{2 \pi \alpha}{ L_x} , \frac{2\pi \beta }{ L_y} , \frac{2\pi \gamma }{ L_z }\right) \nonumber
\\ &= V_\text{pix} e^{-i \vec k \cdot \vec x_\text{min} } F_{\alpha \beta \gamma} .
\label{eq:fftABC}
\end{align}

Returning to the specific example of the momentum form factor $f_s(\vec q)$:
\begin{align}
f_s^\star( \vec q) &= \mathcal F[\Phi_{1 \rightarrow 2}^\star]
\simeq V_\text{pix} e^{+i \vec q \cdot \vec x_\text{min} } F_{\alpha \beta \gamma}^\star \nonumber\\
&= V_\text{pix} \exp\left( + i \vec q \cdot \vec x_\text{min} \right) F_{\alpha \beta \gamma}^\star ,
\end{align}
where 
\begin{align} \label{eq:qabc}
\vec q &= \frac{2\pi \alpha }{L_x} \hat x + \frac{2 \pi \beta }{L_y} \hat y + \frac{2 \pi \gamma }{L_z} \hat z ,
\end{align}
and where $F^\star_{\alpha \beta \gamma}$ is to be understood as the forward FFT of the complex conjugate array,
\begin{align}
f^\star_{a b c} &= \Phi_{1 \rightarrow 2}^\star(x_a, y_b, z_c) .
\end{align}

\subsection{Practical Considerations}\label{sec:limitations}

The primary benefits of the FFT are its simplicity and easy adaptability: once $\Phi_{g \rightarrow s}(\vec r)$ has been tabulated, it no longer matters which types of basis functions were used to generate it. 
On the other hand, an accurate depiction of $f_s(\vec q)$ requires a position-space grid that is both large in volume and finely sampled in $\vec r$.

From \eqref{eq:qabc}, the ``pixel'' size in the momentum space grid is given by $2\pi / L$, where $L$ is the length of the integration volume. 
Generating a $0.1$ keV grid in $f_s(\vec q)$ would necessitate an $L \approx 124$~angstrom box size, which is two orders of magnitude larger than a typical bond length. Simultaneously, we still need to sample $\Phi_{g \rightarrow s}(\vec r)$ finely enough to capture all of its relevant features, otherwise \eqref{eq:continuFFT} will be inaccurate. 
Even if we set $\Phi_{g \rightarrow s}(\vec r) \rightarrow 0$ for $\vec r \gg \text{few} \times a_0$ far away from the molecule, such a 3d grid can easily exceed the available computer memory.

\begin{figure*}
\centering
\includegraphics[width=\textwidth]{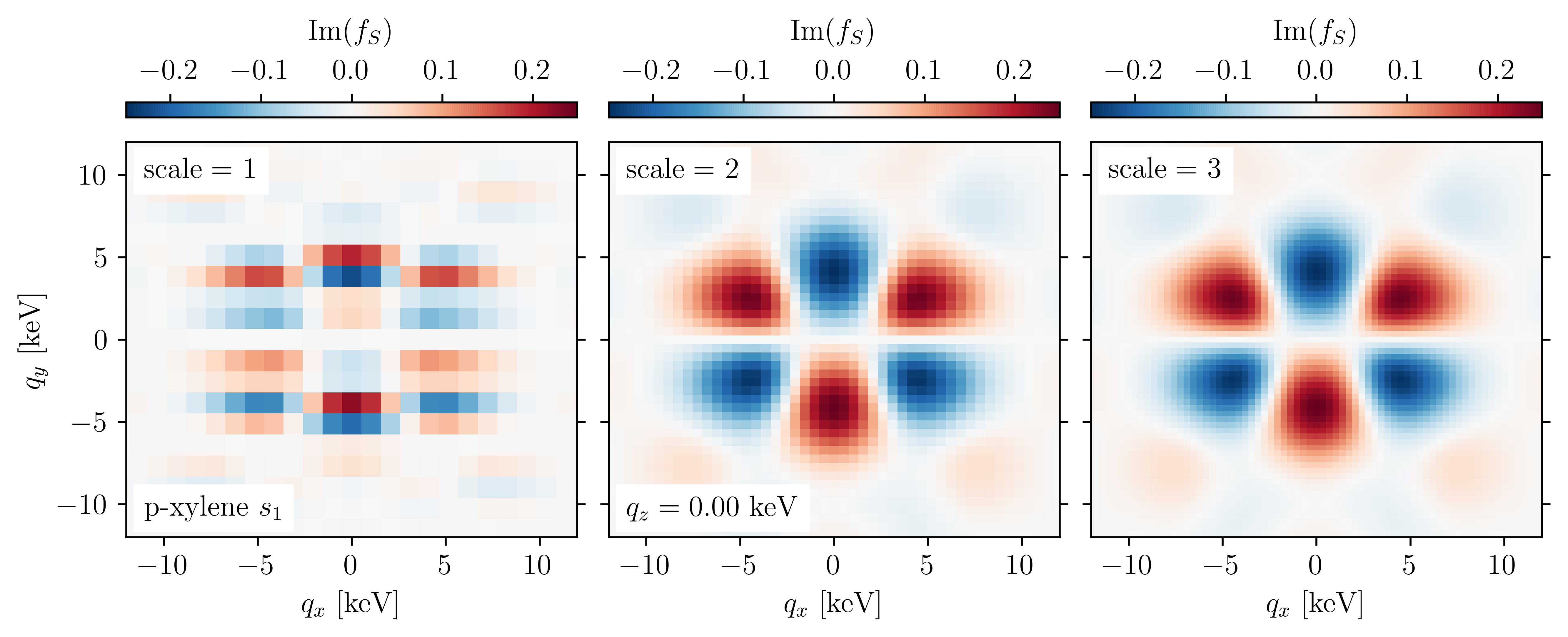}
\caption{The accuracy of the FFT method is controlled in part by the volume of the sampled region. The ``scale = 1'' volume encloses the atomic coordinates in a rectangular box with a $5 a_0$ margin in all directions. Its qualitative accuracy is poor, although it correctly identifies the order of magnitude. In the ``scale = 2'' and ``scale = 3'' examples, we extend the integration volume by a factor of two or three before applying the FFT, yielding more accurate and finely grained pictures of $f_s(\vec q)$, shown here in the $q_z = 0$ plane. 
}
\label{fig:FFTconverge}
\end{figure*}

Even so, the FFT can be highly efficient at generating low-to-medium precision form factors. Figure~\ref{fig:FFTconverge} shows the convergence towards accuracy as the integration volume is expanded, using the first excited state of p-xylene as an example.
Para-xylene is symmetric under central inversion, $\vec r \rightarrow - \vec r$, 
which implies that $f_s = i \,\text{Im}\,f_s$ for any excited states with a nonzero transition dipole moment~\cite{Giffin:2025hdx}. 
Initially, $\Phi_{g \rightarrow s}$ is tabulated within a rectangular box of dimensions $L_{x, y, z} \simeq (12.0, 9.7, 7.4)$ angstrom, which includes a margin of $5 a_0$ in all directions away from the atomic positions. 
This ``scale = 1'' FFT does a poor job of recreating $f_s$, even qualitatively, except for the locations of the $f_s = 0$ nodes. The grid spacing $\Delta x = 0.07$\,angstrom is more than sufficient to capture the small-scale features in $\Phi_{g \rightarrow s}(\vec r)$, so the imprecision is due to $L_{x,y,z}$.

In the second and third panels of Figure~\ref{fig:FFTconverge}, the integration region is extended to $\text{scale} \times (12.0 \,\text{angstrom})$ in every direction, for ``scale = 2'' and ``scale = 3''. The $L \approx 24$\,angstrom box correctly reproduces all of the features in $f_s(\vec q)$, with a pixelation that is apparent to the eye. At $L \approx 36$\,angstrom, $f_s(\vec q)$ resolves into a clearer image, without revealing any substantial differences with the ``scale = 2'' version.
Whether the inaccuracy in $f_s$ will affect the scattering rate in \eqref{eq:rateR} depends largely on the DM model in question: the accuracy is probably sufficient for heavy mediator models, $F_\text{DM} = 1$. For light mediators, $F_\text{DM} = (\alpha m_e)^2/q^2$, the enhancement at small $q \lesssim \alpha m_e$ will magnify the pixelation error.

In conclusion, while the FFT method is simple to implement, we must ensure that the position space and momentum space grids are both sufficiently fine-grained to capture all of the physical effects in the system. A compromise can often be found between accuracy and speed.
In the next section, we introduce a pair of analytic methods which evaluate $f_s(\vec q)$ exactly, at arbitrary values of $\vec q$. By ``exact,'' 
we mean that  we integrate \eqref{eq:fSreal} without any loss of precision.

\section{Analytic Method} \label{sec:analytic}
The Fourier transform method, whilst conceptually simple, suffers from the grid resolution issues discussed in Sec.~\ref{sec:limitations}. As such, \texttt{SCarFFF} also implements methods based on analytic expressions for the form factor, which we will now discuss. 

Our starting point is the expression for the form factor~\eqref{eq:fSreal}, written in the atomic orbital (AO) basis as
\begin{equation}
    f_s(\vec{q)} = 2\sum_{\alpha\beta,IJ} \int d^3r\, \phi^*_{\alpha}(\vec{r}-\vec{r}_I) \phi_{\beta}(\vec{r}-\vec{r}_J) \,T^{(s)}_{\alpha\beta} \,e^{i\vec{q}\cdot \vec{r}},
\end{equation}
with $T_{\alpha\beta}^{(s)}$ the transition density matrix for singlet state $s$ in the AO basis, and $\phi_\alpha$ the GTO corresponding to orbital $\alpha$, defined in Appendix~\ref{appx:basis}. Each of these orbitals can be expanded in terms of sums of primitives, which can, in turn, be expanded in terms of sums of their individual Cartesian polynomial terms. The indices $I$ and $J$ refer to individual atoms within the molecule, located at positions $\vec{r}_I$ and $\vec{r}_J$, respectively.

To simplify notation and avoid confusion, we note that each Cartesian term belongs a primitive, which in turn belongs an orbital and atom within the molecule. We will therefore express everything in terms of the Cartesian term indices $i$ and $j$ in what follows, so that \textit{e.g.} $T^{(s)}_{ij}$ corresponds to the component of $T^{(s)}_{\alpha\beta}$ implied by Cartesian terms $i$ and $j$, and $\vec{r}_{i}$ similarly refers to the position $r_\vec{I}$. This reduces the outer sum to a simpler one over $i$ and $j$. 

Expanding out the GTOs, we therefore find
\begin{align}
    f_s(\vec{q}) &= 2\sum_{i,j} T^{(s)}_{ij}\, C_{ij} \int_{-\infty}^{\infty} dx\, (x-x_i)^{a_i} (x-x_{j})^{a_j} \\
    &\times \exp\left(i q_x - \frac{(x-x_i)^2}{2\sigma_i^2} - \frac{(x-x_i)^2}{2\sigma_i^2}\right) \int_{-\infty}^{\infty} dy\dots, \notag
\end{align}
where $C_{ij}$ contains all of the primitive scale factors, normalization coefficients, and Cartesian term weights for this $ij$ pair, and where $a_i$ and $a_j$ are the exponents of the Cartesian terms, \textit{e.g.} $a = 1$ for the $p_x$-orbital, or $a = 0$ for an $s$-orbital. We define $b_i$ and $c_i$ similarly for the $y$- and $z$-components of an orbital. 

Each of these one-dimensional integrals can be solved by first completing the square on the polynomial appearing in the exponent, followed by a redefinition of the integral measure to leave
\begin{align}
    f_s(\vec{q}) &= 2\sum_{i,j} T^{(s)}_{ij} C_{ij} \exp\left(-\frac{r_{ij}^2}{2\bar{\sigma}_{ij}^2} + i\vec{q}\cdot \vec{R}_{ij}\right)\\&\times\int_{-\infty}^{\infty}dx\, (x+ \mathcal{X}_{ij}-x_i)^{a_i}(x+\mathcal{X}_{ij}-x_j)^{a_j}\notag \\
    &\times \exp\left(iq_x x - \frac{x^2}{2\sigma_{ij}^2}\right)\int_{-\infty}^{\infty} dy\dots,\notag
\end{align}
where we have defined
\begin{equation}\label{eq:sigmaTidbits}
    \sigma_{ij} = \frac{\sigma_i \sigma_j}{\bar{\sigma}_{ij}}, \quad \bar{\sigma}_{ij} = \sqrt{\sigma_i^2 + \sigma_j^2},
\end{equation}
and
\begin{equation}\label{eq:positionTidbits}
    \mathcal{X}_{ij} = \sigma_{ij}^2 \left(\frac{x_i}{\sigma_i^2} + \frac{x_j}{\sigma_j^2}\right), \quad x_{ij} = |x_i - x_j|,
\end{equation}
with similar expressions for $\mathcal{Y}_{ij}$ and $\mathcal{Z}_{ij}$.   $\vec{R}_{ij}$ is simply the vector $(\mathcal{X}_{ij},\mathcal{Y}_{ij},\mathcal{Z}_{ij})$. To further facilitate the integral, the polynomials can be conveniently rewritten using the binomial theorem as 
\begin{equation}
    (x+ \mathcal{X}_{ij}-x_i)^{a_i} (x+ \mathcal{X}_{ij}-x_j)^{a_j} = \sum_{A=0}^{\alpha_{ij}} b_{ij}^{A} x^A,
\end{equation}
with $\alpha_{ij} = a_i + a_j$, and the $b$ coefficients defined by
\begin{align}\label{eq:bCoefficients}
    b_{ij}^A &= \sum_{k = \mathrm{max}\{0,\, a_i - A\}}^{\mathrm{min}\{a_i, \,\alpha_{ij} - A\}} \binom{a_i}{k} \binom{a_j}{\alpha_{ij} - k - A} \\
    &\times(\mathcal{X}_{ij} - x_i)^k (\mathcal{X}_{ij} - x_j)^{\alpha_{ij} - k - A},\notag
\end{align}
where once again, we define similar $b$ coefficients for the $y$- and $z$- directions indexed by $B$ and $C$. This allows us to write each of the integrals in terms of differential operators, and perform the resulting Gaussian integral to leave
\begin{align}
    f_s(\vec{q}) &= 2(2\pi)^\frac{3}{2}\sum_{i,j} T^{(s)}_{ij}M_{ij}  ,e^{i\vec{q}\cdot \vec{R}_{ij}}\\
    &\times \sum_{A=0}^{\alpha_{ij}} b_{ij}^A (-i\partial_{q_x})^A  \exp\left(-\frac{q_x^2 \sigma_{ij}^2}{2}\right) \sum_{B=0}^{\beta_{ij}}\dots,\notag
\end{align}
where concretely,
\begin{equation}\label{eq:Mij}
    M_{ij} = d'_i d'_j k_{i} k_{j} \sigma_{ij}^3 \exp\left(-\frac{r_{ij}^2}{2\bar{\sigma}_{ij}^2}\right),
\end{equation}
where $d'_i = d_i \xi_i N_i$, with the orbital normalisation factors $\xi$, $N$, and $d$ defined in Appendix~\ref{appx:basis}.
Finally, noticing that the derivative term can be rewritten in terms of the \textit{probabilist's} Hermite polynomials
\begin{align}
    (-i\partial_{q_x})^A \exp\left(-\frac{q_x^2\sigma_{ij}^2}{2}\right) &= (i\sigma_{ij})^{A} \mathrm{He}_A(q_x \sigma_{ij}) \\
    &\times \exp\left(-\frac{q_x^2\sigma_{ij}^2}{2}\right)\notag,
\end{align}
we arrive at a useful expression for the form factor, written in terms of three tensors, each depending on just one of the three Cartesian momentum coordinates
\begin{align}\label{eq:CartesianMethod}
    f_s(\vec{q}) = 2(2\pi)^\frac{3}{2} \sum_{i,j} T^{(s)}_{ij} M_{ij} \mathcal{V}^{(x)}_{ij}(q_x) \mathcal{V}^{(y)}_{ij}(q_y)  \mathcal{V}^{(z)}_{ij}(q_z),
\end{align}
where each $\mathcal{V}$-tensor is defined by
\begin{align}\label{eq:CartesianVector}
    \mathcal{V}^{(x)}_{ij}(q_x) &= \exp\left(iq_x \mathcal{X}_{ij} - \frac{q_x^2 \sigma_{ij}^2}{2}\right)\sum_{A=0}^{\alpha_{ij}} b_{ij}^A (i\sigma_{ij})^A \\
    &\times \mathrm{He}_A(q_x \sigma_{ij})\notag.
\end{align}
For those more inclined to use the \textit{physicist's} Hermite polynomials, these are related by
\begin{equation}
    \mathrm{He}_n(x) = 2^{-\frac{n}{2}} H_{n}\left(\frac{x}{\sqrt{2}}\right).
\end{equation}
The expression~\eqref{eq:CartesianMethod} is particularly useful for evaluating the form factor, as each of the $\mathcal{V}$-tensors is a sum of simple polynomials in each of the momenta. Additionally, as we will discuss at length in Sec.~\ref{sec:CartesianMethod}, being able to decompose the form factor into the three momentum directions can be particularly useful when evaluating certain kinds of integrals involving the form factor. We therefore implement~\eqref{eq:CartesianMethod} as one of the methods in \texttt{SCarFFF}. 

We now note that~\eqref{eq:CartesianMethod} is not a the sole useful analytic expression for the form factor. For several reasons, that we will discuss in Sec.~\ref{sec:SphericalMethod}, it is often more convenient and computationally efficient to compute the form factor in spherical coordinates. To do so, we first rewrite the product of Hermite polynomials contained in each of the $\mathcal{V}$-tensors, packaged alongside several of the other coefficients, in terms of a sum of monomials, to arrive at
\begin{align}
    f_s(\vec{q}) &= 2(2\pi)^{\frac{3}{2}} \sum_{i,j} T^{(s)}_{ij} \exp\left(-\frac{q^2 \sigma_{ij}^2}{2}\right) e^{i\vec{q}\cdot \vec{R}_{ij}} \\
    &\times \sum_{u,v,w} \mathcal{D}^{uvw}_{ij} q_x^u q_y^v q_z^w, \notag
\end{align}
where $u$ runs from $0$ to $\alpha_{ij}$, and $v$ and $w$ follow the same pattern for the $y$- and $z$- directions, whilst the $\mathcal{D}$-tensor is defined by
\begin{equation}\label{eq:DTensor}
    \mathcal{D}_{ij}^{uvw} = M_{ij} \mathcal{C}_{ij}^u \mathcal{C}_{ij}^v \mathcal{C}_{ij}^w,
\end{equation}
with
\begin{equation}\label{eq:Ccoefficients}
    \mathcal{C}_{ij}^u = i^u \sum_{m=0}^{\left\lfloor \frac{\alpha_{ij} - u}{2} \right\rfloor} b_{ij}^{u+2m}\sigma_{ij}^{2(u+m)} \frac{(u+2m)!}{m!u!\,2^m},
\end{equation}
and similar in $v$ and $w$ for the $y$- and $z$- components. The next step is somewhat more complicated, and involves projecting the monomials onto a basis of spherical harmonics. As the derivation is rather opaque, we leave this in Appendix~\ref{sec:thetaCoefficients} for the more interested reader, and simply quote the resulting form factor here:
\begin{align}
    f_s(\vec{q}) &= 2(2\pi)^{\frac{3}{2}} \sum_{i,j} T^{(s)}_{ij} \exp\left(-\frac{q^2 \sigma_{ij}^2}{2}\right) e^{i\vec{q}\cdot \vec{R}_{ij}} \\
    &\times \sum_{u,v,w} \mathcal{D}^{uvw}_{ij} q^n \sum_{\lambda \in \{n, n-2, \dots\}}\sum_{\mu = -\lambda}^\lambda \Theta_{\lambda\mu}^{uvw} Y_{\lambda}^{\mu}(\hat q), \notag
\end{align}
with the $\Theta_{\lambda\mu}^{uvw}$ coefficients also defined in Appendix~\ref{sec:thetaCoefficients}, $n = u + v + w$, and where $Y_{\lambda}^{\mu}$ denotes a \textit{complex} spherical harmonic, with the Condon-Shortley phase included. After a careful reindexing, and reordering of the summation order, this expression can be simplified to
\begin{align}
    f_s(\vec{q}) &= 2(2\pi)^{\frac{3}{2}} \sum_{i,j} T^{(s)}_{ij} \exp\left(-\frac{q^2 \sigma_{ij}^2}{2}\right) e^{i\vec{q}\cdot \vec{R}_{ij}} \\
    &\times \sum_{n} q^n \sum_{\lambda \in \{n, n-2, \dots\}}\sum_{\mu = -\lambda}^\lambda \mathcal{W}_{ij,\lambda\mu}^{n} Y_{\lambda}^{\mu}(\hat q), \notag
\end{align}
where 
\begin{equation}\label{eq:WTensor}
    \mathcal{W}_{ij,\lambda\mu}^{n} = \sum_{\substack{u,v,w \\ u+v+w =n}}\mathcal{D}_{ij}^{uvw}\Theta_{\lambda\mu}^{uvw}.
\end{equation}
Owing to its symmetries and selection rules, explicitly its invariance under $i \leftrightarrow j$, along with the conditions $n \geq \lambda \geq |\mu|$, $|\mu| \leq u + v$, and that $n$ and $\lambda$ must share the same parity, the $\mathcal{W}$-tensor turns out to be a particularly efficient object to compute, and forms the basis for the spherical grid method implemented in \texttt{SCarFFF}. However, we have yet to completely disentangle the radial and angular components of the form factor. The next step in this process is to perform a plane-wave expansion of the phase factor
\begin{equation}
    e^{i\vec{q}\cdot \vec{R}_{ij}} = 4\pi \sum_{L,M} i^L j_L(qR_{ij}) Y_{L}^M(\hat{q}) Y_{L}^{M*}(\hat{R}_{ij}),
\end{equation}
with $j_L$ the spherical Bessel function of order $L$. The spherical harmonic that depends on $\hat q$ resulting from this expansion can be coupled to existing one from the monomial-to-spherical transformation a Gaunt coefficient, as
\begin{equation}
    Y_{\lambda}^\mu(\hat{q})Y_{L}^M(\hat{q}) = \sum_{\ell}\mathcal{G}_{\lambda \mu, LM}^{\ell m} Y_{\ell}^m(\hat{q})
\label{eq:YYY}
\end{equation}
with the Gaunt coefficient defined by 
\begin{align}
    \mathcal{G}_{\lambda \mu, LM}^{\ell m} &= (-1)^m \sqrt{\frac{(2\lambda+1)(2L+1)(2\ell+1)}{4\pi}}\\
    &\times\tj{\lambda}{L}{\ell}{\mu}{M}{-m}\tj{\lambda}{L}{\ell}{0}{0}{0},\notag
\end{align}
where the large bracketed objects are Wigner-$3j$ symbols, whose selection rules fix $M = m - \mu$ hereafter. We note that the Gaunt coefficients can also be written as the integral of a product of three spherical harmonics, but that $\mathcal{G}_{\lambda \mu, LM}^{\ell m}$ is more efficiently calculated from the Wigner-$3j$ symbols. 

This leaves us with our final expression for the form factor,
\begin{equation}\label{eq:SphericalMethod}
    f_s(\vec{q}) =  \sum_{\ell,m} Y_{\ell}^m(\hat{q}) \mathcal{R}_{\ell m}(q),
\end{equation}
which conveniently factorizes the angular dependence from the $|\vec q|$-dependent function $\mathcal R_{\ell m}(q)$.
All of the difficult analytic structure is contained within the $\mathcal R$ tensor, 
\begin{align}\label{eq:RTensor}
    \mathcal{R}_{\ell m}(q) &= 4(2\pi)^\frac{5}{2} \sum_{i,j} \exp\left(-\frac{q^2 \sigma_{ij}^2}{2}\right) \sum_{L} i^L \\
    &\times j_L(qR_{ij}) Y_{L}^{M*}(\hat{R}_{ij}) \sum_{n} q^{n} \sum_{\lambda,\mu} \mathcal{W}_{ij,\lambda\mu}^{n}\,\mathcal{G}_{\lambda \mu, LM}^{\ell m},\notag
\end{align}
where we remind the reader that $M = m - \mu$. 

\begin{figure*}
\centering
\includegraphics[width=0.51\textwidth]{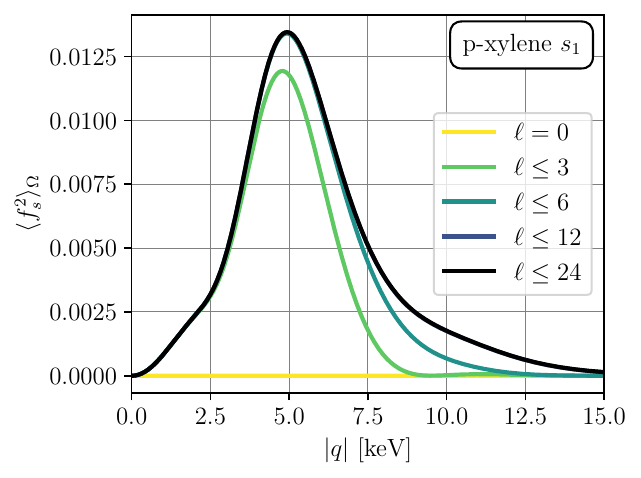}
\hspace{-2em}
\includegraphics[width=0.5\textwidth]{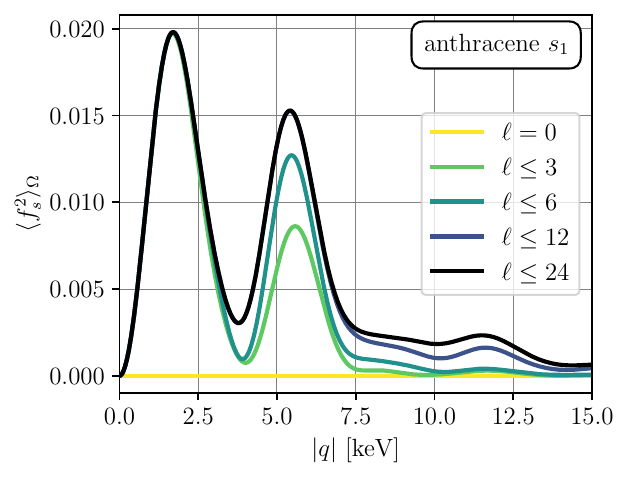}
\caption{
We show the isotropic average form factor $(4\pi)^{-1} \int d\Omega  \, |f_s|^2$ for the first excited states of p-xylene and anthracene, calculated with the spherical analytic method of Section~\ref{sec:SphericalMethod}.
This $f_s$ is found by a sum over spherical harmonic modes $f_s(\vec q) = \sum_{\ell m} Y_l^m(\hat q) \mathcal R_{lm}(q)$, which we truncate at finite $\ell$.
We find that the first excited state of p-xylene is very well described by the $\ell \leq 12$ angular modes, while an accurate representation of anthracene above $q > 7$ keV requires the $\ell > 12$ modes.
}
\label{fig:ellConverge}
\end{figure*}

The sum over $\ell$ formally runs from $0$ to $\infty$, but we show in Figure~\ref{fig:ellConverge}  that this sum can be safely truncated at finite $\ell_\mathrm{max}$.
For relatively simple molecules such as benzene ($C_6 H_{6}$) and p-xylene ($C_8 H_{10}$), the sum over $\ell$ converges by $\ell_\text{max} = 12$.
Larger molecules, e.g.~anthracene ($C_{14} H_{10}$), may require $\ell_\text{max}$ in the range of 18--24, particularly at high momenta ($q \gg \alpha m_e$). 
As a 1d proxy for the 3d form factor, Figure~\ref{fig:ellConverge} shows the isotropic angular average, defined 
\begin{align}
\langle f_s^2 \rangle_\Omega &\equiv \int \! \frac{d\Omega }{4 \pi} \left| f_s(\vec q) \right|^2 .
\end{align}
From the orthogonality properties of spherical harmonics, it can be shown that 
\begin{align}
\langle f_s^2 \rangle_\Omega & = \frac{1}{4 \pi } \sum_{\ell m} \left| \mathcal R_{\ell m}(q) \right|^2 .
\end{align}
In an isotropic medium such as a fluid or glass, the scattering rate depends only on $\langle f_s^2(\vec q) \rangle_\Omega$, so it is quite convenient that it can be extracted directly from the $\mathcal R_{\ell m}^2$ sum. 

\subsection{Summary}

We implement~\eqref{eq:SphericalMethod} within \texttt{SCarFFF} to tabulate the form factor on a uniform spherical grid. This has advantages over the other two methods, as it is far more robust against numerical issues than the FFT method, but far faster to materialise the full 3d form factor grid than the Cartesian method, at the expense of a more difficult derivation. We will now move on to discuss the numerical implementations of each method in detail.

\section{Molecular Form Factors with \texttt{SCarFFF}}
\begin{figure*}[p]
  \centering
  \makebox[\textwidth][c]{%
  \begin{tikzpicture}[
    x=\textwidth, y=-0.88\textheight,
    font=\small,
    proc/.style={ 
      draw,
      rounded corners=8pt,
      fill=black!5,
      align=center,
      minimum width=0.17\textwidth,
      minimum height=0.06\textheight
    },
    procCol/.style={ 
      draw,
      rounded corners=8pt,
      fill=black!5,
      align=center,
      minimum width=0.17\textwidth,
      minimum height=0.05\textheight
    },
    arr/.style={-Stealth, thick}, 
    seg/.style={thick},           
    setbox/.style={
      rounded corners=14pt,
      thick,
      inner xsep=10pt,
      inner ysep=18pt
    },
    output/.style={fill=cyan!18},
    setboxGeom/.style  ={setbox, draw=violet!60!black, fill=violet!6, inner xsep=14pt},
    setboxDFT/.style   ={setbox, draw=purple!60!black, fill=purple!6, inner xsep=14pt},
    setboxSecond/.style={setbox, draw=orange!70!black, fill=orange!8},
    setboxCol1/.style  ={setbox, draw=teal!60!black,   fill=teal!7},
    setboxCol2/.style  ={setbox, draw=red!60!black,    fill=red!6},
    setboxCol3/.style  ={setbox, draw=blue!60!black,   fill=blue!6},
    settitle/.style={font=\bfseries, align=center, text width=0.32\textwidth},
    settitleNowrap/.style={font=\bfseries, align=center}
  ]
  \coordinate (NW) at (0,0);

  \def\rowOneY{0.08}
  \def\rowGap{0.2}
  \def\rowTwoY{\rowOneY+\rowGap}

  \node[proc] (A1) at ($(NW)+(0.10,\rowOneY)$) {SMILES\\Basis Set};
  \node[proc] (A2) at ($(NW)+(0.35,\rowOneY)$) {\texttt{RDKit} ETKDGv3\\ + UFF};
  \node[proc] (A3) at ($(NW)+(0.55,\rowOneY)$) {\texttt{PySCF} + \texttt{geomeTRIC}};
  \node[proc] (A4) at ($(NW)+(0.82,\rowOneY)$) {\texttt{PySCF}};

  \node[proc] (B1) at ($(NW)+(0.10,\rowTwoY)$) {Read TD-DFT\\+ geometry};
  \node[proc] (B2) at ($(NW)+(0.35,\rowTwoY)$) {Read GTO\\parameters};
  \node[proc] (B3) at ($(NW)+(0.60,\rowTwoY)$) {Build hierarchical\\1D arrays};
  \node[proc] (B4) at ($(NW)+(0.85,\rowTwoY)$) {Build index maps:\\Cart$\to$prim$\to$orb$\to$atom};

  \begin{scope}[on background layer]
    \node[setboxGeom]   (GeomBox)     [fit=(A2)(A3)] {};
    \node[setboxDFT]    (DFTBox)      [fit=(A4)] {};
    \node[setboxSecond] (SecondRowBox)[fit=(B1)(B2)(B3)(B4)] {};

    \node[settitleNowrap, text=violet!60!black, anchor=north]
      at ([yshift=-3pt]GeomBox.north) {Geometry Optimisation};
    \node[settitleNowrap, text=purple!60!black, anchor=north]
      at ([yshift=-3pt]DFTBox.north) {DFT and TD-DFT};
    \node[settitleNowrap, text=orange!70!black, anchor=north]
      at ([yshift=-3pt]SecondRowBox.north) {Molecule Construction};
  \end{scope}

  \def\colDx{0.30}
  \def\colFirst{0.15}
  \def\colStep{0.08}

  \coordinate (Row2Anchor) at (SecondRowBox.south);
  \coordinate (Col1Anchor) at ($(Row2Anchor)+(-\colDx,0)$);
  \coordinate (Col2Anchor) at ($(Row2Anchor)+(0,0)$);
  \coordinate (Col3Anchor) at ($(Row2Anchor)+(\colDx,0)$);

  \node[procCol]        (C1a) at ($(Col1Anchor)+(0,\colFirst)$)                 {Compute pair arrays \\ $\sigma_{ij},\bar{\sigma}_{ij},M_{ij}, \vec{R}_{ij}, r_{ij}$};
  \node[procCol]        (C1b) at ($(Col1Anchor)+(0,\colFirst+\colStep)$)        {Compute $b$ coefficients\\ $b_{ij}^A, b_{ij}^B, b_{ij}^C$};
  \node[procCol]        (C1c) at ($(Col1Anchor)+(0,\colFirst+2*\colStep)$)      {Compute $\mathcal{D} $ tensor\\$\mathcal{D}_{ij}^{uvw}$};
  \node[procCol]        (C1d) at ($(Col1Anchor)+(0,\colFirst+3*\colStep)$)      {Read $\Theta$ and Gaunt\\$\Theta_{\lambda\mu}^{uvw}, \mathcal{G}_{\lambda L \ell}^{\mu m}$};
  \node[procCol]        (C1e) at ($(Col1Anchor)+(0,\colFirst+4*\colStep)$)      {Compute $\mathcal{W}$ tensor\\$\mathcal{W}_{ij,\lambda \mu}^{n}$};
  \node[procCol,output] (C1f) at ($(Col1Anchor)+(0,\colFirst+5*\colStep)$)      {Compute $\mathcal{R}$ tensor\\ $\mathcal{R}_{\ell m}(q)$};
  \node[procCol,output] (C1g) at ($(Col1Anchor)+(0,\colFirst+6*\colStep)$)      {Contract to full grid\\$f_s(\vec{q})$};

  \node[procCol]        (C2a) at ($(Col2Anchor)+(0,\colFirst)$)                 {Compute pair arrays \\ $\sigma_{ij},\bar{\sigma}_{ij},M_{ij}, \vec{R}_{ij}, r_{ij}$};
  \node[procCol]        (C2b) at ($(Col2Anchor)+(0,\colFirst+\colStep)$)        {Compute $b$ coefficients\\ $b_{ij}^A, b_{ij}^B, b_{ij}^C$};
  \node[procCol,output] (C2c) at ($(Col2Anchor)+(0,\colFirst+2*\colStep)$)      {Compute $\mathcal{V}$ tensors\\$\mathcal{V}_{ij}(q_x),\mathcal{V}_{ij}(q_y),\mathcal{V}_{ij}(q_z)$};
  \node[procCol,output] (C2d) at ($(Col2Anchor)+(0,\colFirst+3*\colStep)$)      {Contract to full grid\\$f_s(\vec{q})$};

  \node[procCol,output] (C3a) at ($(Col3Anchor)+(0,\colFirst)$)                 {Compute density\\ $\Phi_{0\to i}(\vec{x})$};
  \node[procCol]        (C3b) at ($(Col3Anchor)+(0,\colFirst+\colStep)$)        {Perform FFT\\$F(\vec{q})$};
  \node[procCol,output] (C3c) at ($(Col3Anchor)+(0,\colFirst+2*\colStep)$)      {Normalise, rephase\\ $f_s(\vec{q})$};

  \node[procCol,output] (PlotBox)   at ($(Col3Anchor)+(0,\colFirst+5*\colStep)$) {Plotting};
  \node[procCol] (VerifyBox) at ($(Col3Anchor)+(0,\colFirst+6*\colStep)$) {Verification};

  \begin{scope}[on background layer]
    \node[setboxCol1] (Col1Box) [fit=(C1a)(C1b)(C1c)(C1d)(C1e)(C1f)(C1g)] {};
    \node[setboxCol2] (Col2Box) [fit=(C2a)(C2b)(C2c)(C2d)] {};
    \node[setboxCol3] (Col3Box) [fit=(C3a)(C3b)(C3c)] {};

    \node[settitleNowrap, text=teal!60!black, anchor=north]
      at ([yshift=-3pt]Col1Box.north) {Spherical Method};
    \node[settitleNowrap, text=red!60!black, anchor=north]
      at ([yshift=-3pt]Col2Box.north) {Cartesian Method};
    \node[settitleNowrap, text=blue!60!black, anchor=north]
      at ([yshift=-3pt]Col3Box.north) {FFT Method};
  \end{scope}

  \draw[arr] (A2) -- (A3);

  \draw[arr] (B1) -- (B2);
  \draw[arr] (B2) -- (B3);
  \draw[arr] (B3) -- (B4);

  \draw[arr] (C1a) -- (C1b);
  \draw[arr] (C1b) -- (C1c);
  \draw[arr] (C1c) -- (C1d);
  \draw[arr] (C1d) -- (C1e);
  \draw[arr] (C1e) -- (C1f);
  \draw[arr] (C1f) -- (C1g);

  \draw[arr] (C2a) -- (C2b);
  \draw[arr] (C2b) -- (C2c);
  \draw[arr] (C2c) -- (C2d);

  \draw[arr] (C3a) -- (C3b);
  \draw[arr] (C3b) -- (C3c);

  \draw[arr] (A1.east) -- (GeomBox.west);
  \draw[arr] (GeomBox.east) -- (DFTBox.west);

  \coordinate (K) at ($(DFTBox.south)+(0,0.03)$);
  \draw[arr] (DFTBox.south) -- (K) -| (SecondRowBox.north);

  \coordinate (J)  at ($(SecondRowBox.south)+(0,0.04)$);
  \coordinate (Jw) at ($(J)+(-\colDx,0)$);
  \coordinate (Je) at ($(J)+(\colDx,0)$);

  \draw[arr] (SecondRowBox.south) -- (Col2Box.north);
  \draw[arr] (J) -- (Jw) -- (Col1Box.north);
  \draw[arr] (J) -- (Je) -- (Col3Box.north);

  \coordinate (Sgap) at ($(C1f)!0.5!(C1g)$);
  \coordinate (Sout) at ($(Col1Box.east |- Sgap)$);

  \def\splitLeft{0} 

  \coordinate (MergeX) at ($(Col2Box.center)!0.5!(Col3Box.center)$);
  \coordinate (Merge)  at ($(MergeX |- Sgap) + (\splitLeft,0)$);

  \coordinate (Join) at ($(Merge)+(-0.06,0)$);

  \draw[seg] (Sout) -- (Join);
  \draw[seg] (Col2Box.south) |- (Join);

  \coordinate (F1) at ($(Col3Box.south)+(0,0.10)$);
  \coordinate (F2) at ($(Join |- F1)$);
  \draw[seg] (Col3Box.south) -- (F1) -- (F2) -- (Join);

  \draw[seg] (Join) -- (Merge);

  \coordinate (Mplot) at ($(Merge |- PlotBox.west)$);
  \coordinate (Mver)  at ($(Merge |- VerifyBox.west)$);

  \draw[seg] (Merge) -- (Mplot);
  \draw[seg] (Merge) -- (Mver);

  \draw[arr] (Mplot) -- (PlotBox.west);
  \draw[arr] (Mver)  -- (VerifyBox.west);

  \end{tikzpicture}%
  }

  \caption{Workflow of \texttt{SCarFFF}. Given a SMILES string and a basis set set, we first optimize the geometry and perform the DFT and TD-DFT computations. We then build data structures defining the molecule, and pass it through one of the three methods. See the text for more details. Boxes highlighted in blue represent outputs of the code.}
  \label{fig:pipeline}
\end{figure*}
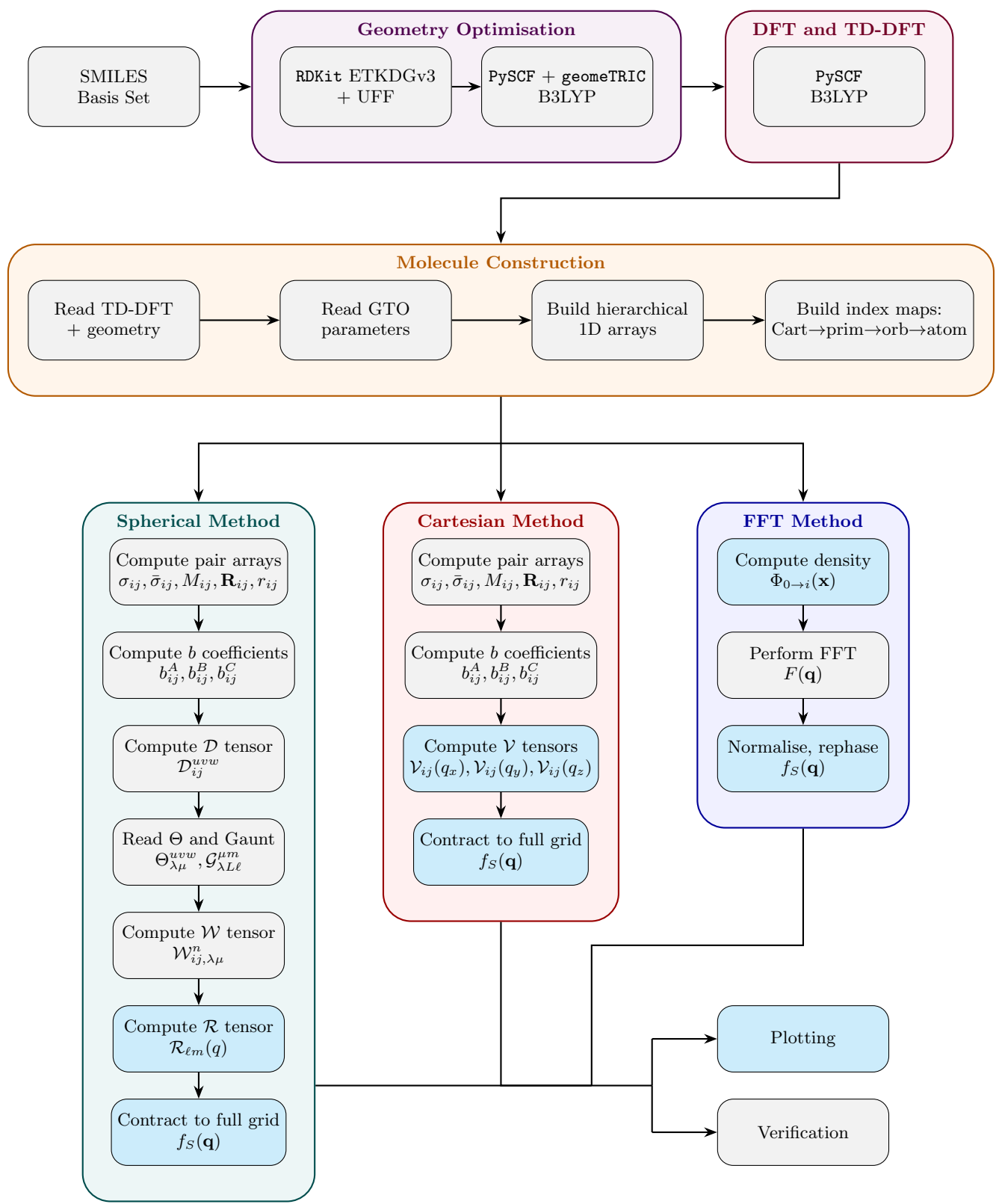

Our numerical package, \texttt{SCarFFF}, implements three methods for the fast and precise computation of molecular form factors: tabulation on a spherical grid, on a Cartesian grid, and using the fast Fourier transform (FFT). Here we describe the numerical implementation of each method, and its advantages and disadvantages with respect to the other methods. For the impatient, we give an \textit{at-a-glance} overview of the numerical package in Fig.~\ref{fig:pipeline}.

\subsection{Molecular Geometry, DFT, and TD-DFT}

 All three methods begin with a computation of the molecular geometry, followed by a ground-state DFT calculation. The resulting ground-state is then used as an input to a TD-DFT calculation of the excited-state transition energies and transition density matrices. 
 
 The molecular geometry optimisation proceeds in several steps. First, we use the \texttt{RDKit}~\cite{rdkit_overview} implementation of the ETKDGv3 algorithm to generate several conformations of the molecule, which we subsequently optimise using the Universal force field (UFF) method. We then take the lowest energy UFF opitmised geometry, and optionally pass this to \texttt{PySCF}~\cite{Sun2020PySCF} and \texttt{geomeTRIC}~\cite{Wang2016geomeTRIC} to optimise further using DFT. This optional last step is significantly slower than the preceding optimisation steps, and offers modest corrections of $\mathcal{O}(0.1\%)$ to the distance matrix entries of small molecules, such as benzene, up to $\mathcal{O}(0.5-3\%)$ for intermediate-sized molecules, such as trans-stilbene, with comparable corrections to the excited state energies in both cases. However, these small corrections to the geometry lead to large corrections to oscillator strength, on average $\mathcal{O}(30\%)$ for the first ten excited states of trans-stilbene, with a maximum of $\mathcal{O}(70\%)$, due to better encoding of molecular symmetries.

To compute the ground- and excited-state energies, we use TD-DFT within \texttt{PySCF} with one of the \texttt{SCarFFF}-supported basis sets. \texttt{SCarFFF} does not depend on the choice exchange-correlation functional, and so in theory any can be used. Throughout this work, we have used B3LYP, although we note that more modern and precise functionals such as \textit{e.g.} $\omega$B97M-V may be more appropriate for certain molecules.  
The excited state energies, atomic coordinates, along with the basis-dependent transition matrices in the atomic orbital basis are then saved to disk in HDF5 format. We also save the computed oscillator strengths for future verification of our form factor results, which we will discuss in Section~\ref{sec:Verification}. The calls to \texttt{PySCF}, including the optional ones to optimise the geometry using DFT, are by far the slowest step in the form factor pipeline. For this reason, we make use of \texttt{GPU4PySCF}, which can be toggled on with an optional \texttt{use\_gpu} flag in the provided scripts. 

\subsection{Fast Fourier Transform Method}\label{sec:FFTMethod}
We now move on to discussing the bulk of the \texttt{SCarFFF} package, beginning with the conceptually simplest approach, the FFT method. This approach directly implements~\eqref{eq:fSreal} as the Fourier transform of the transition density $\Phi^*_{g \to s}(\vec{x})$, with $s$ denoting the $s$-th excited state. 

The first step in the form factor computation is common to all methods. The results from the TD-DFT and geometry optimisation, along with the set of pre-normalised, according to~\eqref{eq:GTOnormalisation}, GTO scale factors and Gaussian widths are read in for the specified basis set. These are used to construct a hierarchical data structure composed of one-dimensional arrays defining the molecule, with index mappings from Cartesian term (\textit{e.g.} the $x^2$ in the $3d_{x^2 -y^2}$ orbital), to GTO primitive, to orbital in a given atom, and finally to atom within the molecule.  This structure saves on both memory and computation time by storing intermediate objects in terms of the largest object in the hierarchy, whilst still allowing us to efficiently sum over the smallest.

Specific to the FFT method, we then construct a spatial grid on which to tabulate the transition density~\eqref{eq:fSreal}. This spatial grid covers all $x \in [-x_\mathrm{lim},x_\mathrm{lim}]$, with resolution $\Delta x$ defined in turn, by
\begin{equation}
    x_\mathrm{lim} = \frac{\pi}{\Delta q_x}, \qquad \Delta x = \frac{2\pi}{q_{x,\mathrm{lim}}},
\end{equation}
with the user-specified momentum grid resolution and maximum, $\Delta q$ and $q_\mathrm{lim}$, respectively. A similar grid is constructed in the $y$- and $z$-directions. Next, the value of the $N_o$ orbitals in the molecule are computed at each grid point, where they are contracted with the full set of transition density matrices to form the transition density function $\Phi_{g \rightarrow s}(\vec r)$. This process features two computationally expensive steps: the tabulation of the orbital values, and the matrix multiplication. These computational costs scale as
\begin{equation}\label{eq:matmulTime}
    c_\mathrm{orb} \simeq k_1 N_p, \qquad c_\mathrm{matmul} \simeq k_2 N_t N_o^2,
\end{equation}
where $N_p$ is the total number of primitives across all orbitals in the molecule, typically around $4 N_o$ for a small basis set such as 6-31g*, $N_t$ is the number of transitions computed per molecule, and $k_1 \simeq 100$ and $k_2 \simeq 2$ some constant prefactors. The cost ratio is therefore approximately
\begin{equation}
    \frac{c_\mathrm{orb}}{c_\mathrm{matmul}} \simeq \frac{200}{N_o N_t},
\end{equation}
which for a typical molecule with a few hundred orbitals scales roughly as $1/N_t$, and is therefore dominated by the matrix multiplication for all but single transition computations. To alleviate this cost, we implement GPU acceleration of the transition density tabulation via custom CUDA kernels, which significantly speeds up the computation at the small expense of a few data transfers. This speedup is particularly noticeable in the matrix multiplication step, which benefits greatly from parallelised GPU operations.

As an additional efficiency measure, we also implement thresholding of small values during the transition density tabulation. This simply sets any GTO primitive whose Gaussian exponent would be sufficiently large, such that the primitive value drops below the threshold value, to zero. This avoids the expensive exponential evaluation that dominates the cost of $k_1$. As an additional measure, if all orbitals at a gridpoint would be zero, we simply skip the expensive matrix multiplication for that point. This last step makes a particularly large difference on spatial grids which extend far beyond the molecule, which as discussed in Section~\ref{sec:limitations} are necessary for good momentum resolution of the form factor.

The final step is to perform the FFT, which is done using the Julia wrapper of \texttt{FFTW}~\cite{FrigoJohnson2005FFTW3}. As this step scales approximately linearly in the total number of gridpoints, it is far faster than the preceding steps. The issue, however, as compared to the analytic methods, is that the FFT step requires the materialisation of both the real space transition grids as well as the momentum space form factor grids. This leads to an additional $50\%$ memory overhead, potentially limiting the size of the grids that can be constructed with this method. The FFT method, whilst fast, can also suffer from numerical artifacts due to insufficient real space resolution and grid extent. A real space grid that is too coarse fails to resolve the peaks at each atomic site. On the other frontier, due to the FFT algorithm assuming periodic functions, a grid sampled only where the molecular wavefunction is large will experience errors due to unphysical \textit{tiling} of the molecule. These can issues can, in turn, be alleviated by increasing $q_{\mathrm{lim}}$ and decreasing $\Delta q$, at a cost in both memory and computation. 

\begin{figure*}
    \centering
    \includegraphics[width=\linewidth]{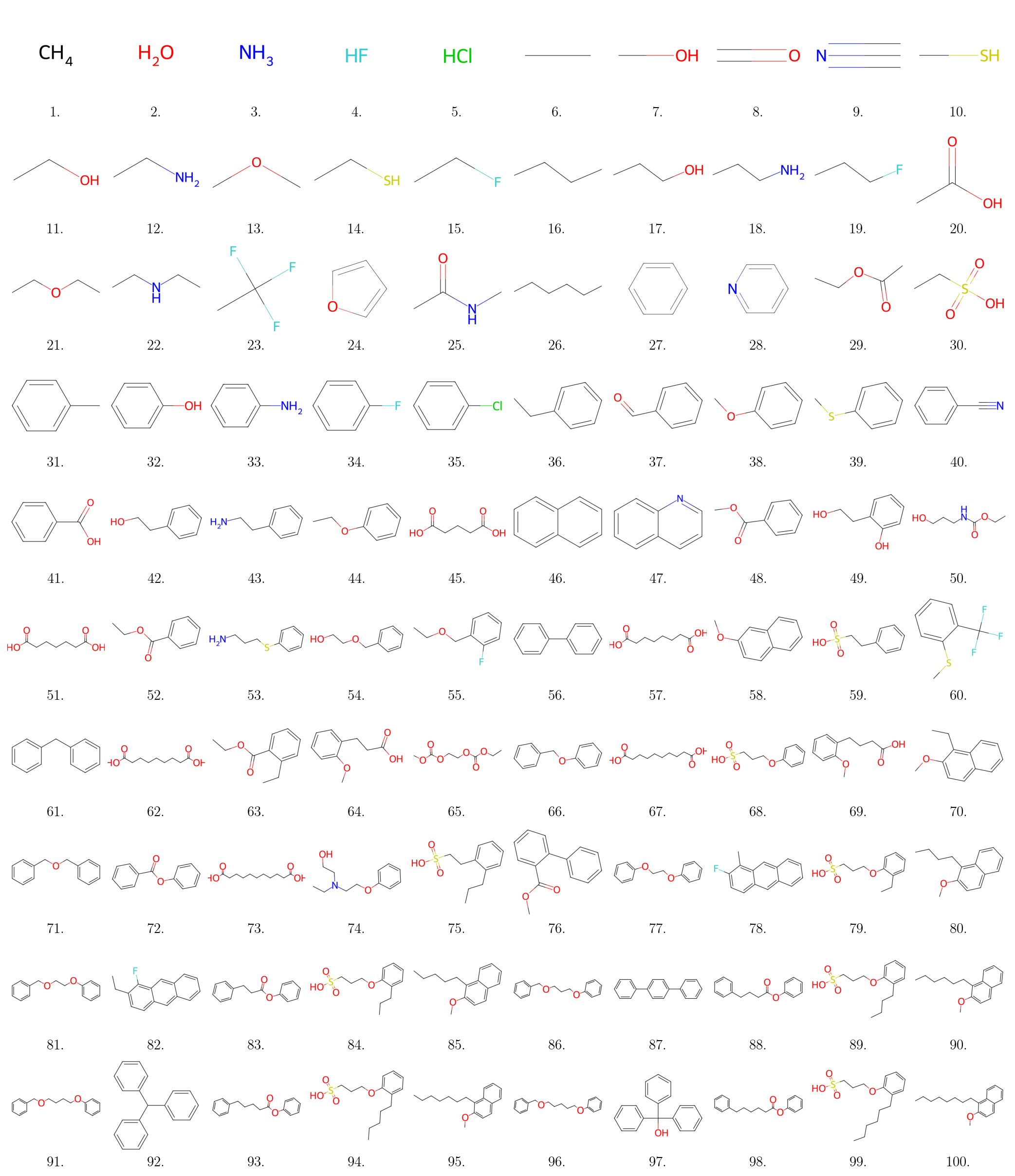}
    \caption{The molecule set used to test \texttt{SCarFFF}. This test set is composed of 100 molecules composed of hydrogen, carbon, nitrogen, sulphur, and fluorine, with an exactly uniformly distributed heavy (non-hydrogen) atom count from 1 to 20. These are chosen to be a combination of long chain and aromatic compounds, so as to represent a wide spectrum of molecules.}
    \label{fig:moltestset}
\end{figure*}

We show the timing of the FFT method on the test molecule set in Fig.~\ref{fig:moltestset}, both with and without GPU acceleration, in Table~\ref{tab:benchmarks}. By comparison with the other two methods, we see that, especially on the CPU path, the FFT method is by far the fastest implementation, and that the GPU speeds up the form factor tabulation by an average factor of $2$, irrespective of grid size. 
It would seem, therefore, that the FFT method is the best choice when running on systems without GPU access, especially when memory is less of an issue, to get a good estimate for molecular form factors. 
We stress, however, that the accuracy of the $100^3$ grid point FFT
is not comparable in accuracy to the same grid using either of the other two methods. For reference, the three plots in Figure~\ref{fig:FFTconverge} used grids of approximately $136^3$, $342^3$, and $512^3$ many points, and only the latter two examples produced approximately correct values of $f_s(\vec q)$. 

For this reason, the FFT method may be best suited for fast searches through large numbers of excited states, where qualitative accuracy is sufficient.
Quantitative accuracy from the FFT generally requires a larger grid, and correspondingly slower evaluation times.
We will discuss the various verification methods implemented in \texttt{SCarFFF} that can be used to alleviate, or at least estimate the magnitude of, these inaccuracies in Sec.~\ref{sec:Verification}.

\begin{table}[t]
\renewcommand{\arraystretch}{1.3}
\renewcommand{\tabcolsep}{9pt}
\begin{tabular}{c|c|c|c|c}
Method                     & $N_{q_i}$ & CPU (s)  & GPU (s) & CPU/GPU      \\
\hline\hline
\multirow{3}{*}{FFT}       & 50        &  $83$   &  $43$  & $1.93\times$ \\
                           & 100       &  $165$       &  $108$        & $1.53\times$             \\
                           & 200       &  $685$        &  $389$       & $2.22\times$              \\
\hline
\multirow{3}{*}{Cartesian} & 50      & $1005$ & $70$      & $9.66\times$     \\
                           & 100       & $7339$        & $166$         &  $44.21\times$            \\
                           & 200       & $48776$        & $520$       &  $93.80\times$            \\
\hline
\multirow{3}{*}{Spherical} & 50        & $1779$ & $334$ & $5.33\times$     \\
                           & 100       &  $3652$        & $611$        & $5.98\times$              \\
                           & 200       &  $8996$        & $1305$       & $6.89\times$             
\end{tabular}
\caption{Run time for the different methods, excluding molecular geometry optimisation and DFT, on the 100 molecule test set, shown in Fig.~\ref{fig:moltestset}, on grids with $N_{q_i}$ points in each direction. For all methods, we use the cc-pVDZ basis set, a threshold of $10^{-6}$, and compute the form factor for the first $12$ excited states up to a maximum of $|\vec{q}| = 15\,\mathrm{keV}$. For the spherical method, we compute up to angular mode $\ell_\mathrm{max} = 24$. For both the pure CPU and GPU accelerated runs, we run \texttt{SCarFFF} on 24 cores of an Intel Xeon Platinum 8468 CPU, whilst the GPU acceleration is performed with an NVIDIA H100 SXM GPU. Not included in these benchmarks is the time for the method-independent geometry optimisation and TD-DFT computations, which took $4736\,\mathrm{s}$ on the same hardware with GPU acceleration.}\label{tab:benchmarks}
\end{table}

\subsection{Cartesian Grid Method}\label{sec:CartesianMethod}
We now turn our discussion to the Cartesian method. This implements~\eqref{eq:CartesianMethod} to tabulate the form factor on a uniform Cartesian grid in $(q_x, q_y, q_z)$. As in Sec.~\ref{sec:FFTMethod}, the first step is to create the hierarchical data structure defining the molecule and its transitions. From here, the Cartesian method diverges from the FFT method by evaluating the Fourier integral analytically, and exploiting the separability in each of the momentum directions.

The next step is precomputing the small matrices of Cartesian pair coefficients defined in~\eqref{eq:Mij},~\eqref{eq:sigmaTidbits},~\eqref{eq:positionTidbits}, and~\eqref{eq:bCoefficients}, which are reused over many computations, and, owing to their small size, can be safely stored in memory. Given these, \texttt{SCarFFF} then constructs the one-dimensional $\mathcal{V}$-tensors defined in~\eqref{eq:CartesianVector}. To speed up the implementation, the Hermite polynomials are generated using recurrence relations up to maximum Cartesian order $\alpha_{ij}$ for each pair. 

The information contained in the $\mathcal{V}$-tensors, along with the small array of pair coefficients $M_{ij}$ and transition density matrices are sufficient to completely define the form factor. This is particularly useful when the form factor is to be used in separable integrals of the form
\begin{equation}
    I_\mathrm{sep} = \int d^3q \,f(q_x) \,g(q_y)\,h(q_z) f_s(\vec{q}),
\end{equation}
where $f,g$ and $h$ are some arbitrary functions each depending on at most on coordinate. Using the separability of the Cartesian form factor, this can be rewritten in the form
\begin{equation}
    I_\mathrm{sep} = \sum_{i,j} T^{(s)}_{ij} M_{ij} \int dq_x \, f(q_x) \mathcal{V}_{ij}(q_x)\int  \dots
\end{equation}
such that the cost of the integration scales linearly with the most densely sampled grid dimension, as opposed to the product of all three in the case of the FFT form factor. Importantly, each of these $\mathcal{V}$-tensors be can constructed in approximately
\begin{equation}
    c_{\mathcal{V}} = k_\mathcal{V} N_c^2 N_{q_i},
\end{equation}
operations, with $i \in \{x,y,z\}$, $N_c \simeq N_p \simeq 4 N_o$ the total number of Cartesian terms, and $k_\mathcal{V} \simeq 300$. Owing to $k_\mathcal{V}$, the construction of the $\mathcal{V}$-tensors will in fact dominate over the integration time, and we can directly compare to the equivalent cost for the FFT method
\begin{equation}
    \frac{c_\mathcal{V}}{c_\mathrm{FFT}} \simeq \frac{2400}{N_{q_i}^2},
\end{equation}
where we assume $N_{q_i}$ is approximately constant for all dimensions, and approximate the time to construct the FFT form factor by the matrix multiplication time~\eqref{eq:matmulTime}. Even for small grids with $N_{q_i} \simeq 100$, this method is therefore far better suited to form factor applications involving separable integrals than the FFT method. To facilitate this application, \texttt{SCarFFF} comes with an option to only compute and save the 1d $\mathcal{V}$-tensors when using the Cartesian method. Much like the FFT method, we also apply thresholding in the Cartesian method by discarding all Cartesian pair index combinations for which $|M_{ij} T^{(s)}_{ij} /(M_{ij} T^{(s)}_{ij})_\mathrm{max}| < \epsilon$ on a transition-by-transition basis, with $\epsilon$ the user-specified threshold parameter. We find that for a threshold of $10^{-6}$, only around 25\% of terms survive for most molecules, speeding up computation by a factor of four  with no measurable impact on the accuracy of the final result. With that said, we highly recommend that the user explores different values of $\epsilon$ for convergence, especially where high accuracy is required. 

Where this method struggles is when the full 3D grid needs to be materialised, as this requires the outer product of the $\mathcal{V}$-tensors, followed by a contraction of the resulting object over the Cartesian term indices. The cost, per gridpoint, of constructing the full form factor (neglecting thresholding) is then
\begin{equation}
    c_\mathrm{Cart} \simeq k_3 N_t N_c^2,
\end{equation}
with $k_3 \simeq 8$. Comparing this to the FFT method, we find
\begin{equation}
    \frac{c_\mathrm{Cart}}{c_\mathrm{FFT}} \simeq 64,
\end{equation}
almost two orders of magnitude slower. However, owing to its simplicity, and robustness to numerical artifacts, this method is particularly useful as a verification tool for the other two methods. This can be done by \textit{e.g.} computing planes of particular interest, or using the 1d $\mathcal{V}$-tensors compute particular points on the full 3d grid. 

To improve the usefulness of this method when the full grid is required, we add a GPU path for the final grid contraction using \texttt{cuBLAS}. As shown in Table~\ref{tab:benchmarks}, this speeds up computation by an extraordinary factor of $\sim 50$ to $100$, depending on the grid size. This makes it highly competitive with the FFT method in terms of speed, without suffering from the same numerical issues. Outside of the GPU path, however, we only recommend that the Cartesian method is only used for verification, or for very small grids with $N_{q_i} \lesssim 50$.

\subsection{Spherical Grid Method}\label{sec:SphericalMethod}
The final method implemented by \texttt{SCarFFF} is the spherical grid method, which tabulates the form factor~\eqref{eq:SphericalMethod} on a uniform grid in $(q, q_\theta,q_\phi)$, where $q = |\vec{q}|$, up to a user-specified $\ell_\mathrm{max}$.

This implementation follows the Cartesian method exactly up to the construction of the Cartesian pair coefficient matrices. From there, we construct the $\mathcal{D}$-tensor defined in~\eqref{eq:DTensor}. Naively, we would need to allocate a full $N_c^2 \times u_\mathrm{max}^3$ object to store the $\mathcal{D}$-tensor, where $u_\mathrm{max}$ is twice the maximum angular momentum in the basis set. However, as many Cartesian pairs have, \textit{e.g.} $\alpha_{ij} < u_\mathrm{max}$, a significant fraction of this dense tensor would be zero. Such a dense tensor would also be inefficient due the symmetry of the $\mathcal{D}$-tensor under $i \leftrightarrow j$. As such, we store the $\mathcal{D}$-tensor in sparse COO format, which for a typical molecule reduces memory requirements by around two orders of magnitude. As an additional benefit, when constructing objects that depend on $\mathcal{D}$, we now only need to iterate over the non-zero entries, which speeds up computation by a comparable amount.  

The next step is to compute the $\Theta$ coefficients defined in Appendix~\ref{sec:thetaCoefficients}, along with the Gaunt coefficients. As these coefficients are independent of the molecule and transition, they are precomputed once for a given $\ell_\mathrm{max}$, and $\lambda_\mathrm{max}$, and then reused across runs. Much like the $\mathcal{D}$-tensor, these would also be incredibly sparse objects if computed in dense form. As such, we exploit the many symmetries and selection rules for these coefficients to compute and store them in sparse COO format.

These are then contracted with the previously computed $\mathcal{D}$-tensor to form the $\mathcal{W}$-tensor defined in~\eqref{eq:WTensor}, which once again, is stored in sparse COO format. Owing to the sparse storage, symmetry exploitation, and selection rules, all steps to this point proceed are completed almost instantly. The next step, however, applying~\eqref{eq:RTensor} to construct the $\mathcal{R}$-tensor, is particularly expensive. We compute the spherical Bessel functions with a fast, custom implementation of Miller's downward recursion algorithm
\begin{equation}
    j_{L-1}(x) = \left(\frac{2L + 1}{x}\right)  j_L(x) - j_{L+1}(x),
\end{equation}
at each $qR_{ij}$ point, giving us access to the full tower of Bessel functions in one sweep. For numerical stability, we always perform this step in float64, and expand compute the Bessel functions using a power series expansion for very small arguments. With this optimisation, the cost of the spherical method is dominated by the accumulation into the $\mathcal{R}$-tensor, and is given by
\begin{equation}
    c_\mathcal{R} \simeq k_\mathcal{R} N_t N_{q} N_\mathcal{W} N_{\bar{\mathcal{G}}},
\end{equation}
where $N_\mathcal{W} \simeq N_c^2$ is the number of non-zero $\mathcal{W}$-tensor entries, $N_{\bar{\mathcal{G}}}$ is the average number of entries for each $(\lambda,\mu)$ combination, and $k_\mathcal{R} \simeq 8$. A rough estimate gives $N_{\bar{\mathcal{G}}} \simeq (\lambda_\mathrm{max} +1)(\ell_\mathrm{max}+1)^2$. Much like the $\mathcal{V}$-tensors in the Cartesian method, the $\mathcal{R}$-tensor contains all of the molecule-specific information about the form factor. As such, it may be useful to terminate the computation early when the form factor is required for integrals where the angular and radial components are separable. We therefore provide an option within \texttt{SCarFFF} to compute only the $\mathcal{R}$-tensor, without materialising the full 3d form factor grid.

Otherwise, the final step is to contract with the angular grid according to~\eqref{eq:SphericalMethod}, the cost for which is dominated by one large outer product, followed by a contraction over $\ell$ and $m$. The cost for this step is roughly
\begin{equation}
    c_\mathrm{contract} \simeq k_4 N_t N_{q} N_{q_\theta} N_{q_\phi} (\ell_\mathrm{max} +1)^2,
\end{equation}
with $k_4 \simeq 2$. Whether this step dominates therefore depends heavily on the size of the angular grid. The cost ratio of this step to that of the $\mathcal{R}$-tensor materialisation is
\begin{equation}
    \frac{c_\mathcal{R}}{c_\mathrm{contract}} \simeq \frac{4 N_c^2 (\lambda_\mathrm{max}+1)}{N_{q_\theta} N_{q_\phi}},
\end{equation}
which for a modest grid with $N_{q_\theta} = N_{q_\phi} \simeq 100$, and a typical molecule with $N_c^2 \simeq 10^5$, $\lambda_\mathrm{max} \simeq 6$, is dominated by the $\mathcal{R}$-tensor construction. Even for a larger grid with $10^3$ points in each direction, we find that the $\mathcal{R}$-tensor step has a comparable cost to the contraction step.

This allows us to directly compare the spherical method to the other two methods as
\begin{equation}
    \frac{c_\mathrm{spher}}{c_\mathrm{FFT}} \simeq \frac{64(\lambda_\mathrm{max}+1)(\ell_\mathrm{max}+1)^2}{N_{q_i}^2},
\end{equation}
and similar for the Cartesian method, with a ratio $64$ times smaller, where we have assumed a similar number of gridpoints in each direction, for each method. For small grids with $N_{q_i} \simeq 100$, and $\ell_\mathrm{max} = 18 - 24$, the spherical method is approximately an order of magnitude slower than the FFT method. However, unlike the FFT method, the spherical method is far less susceptible to numerical errors, and scales far more favourably with the grid size due to the decoupling of the angular and radial parts. 

As with the other two methods, we also offer GPU paths for the spherical method to speed up computation. This includes custom CUDA kernels for the spherical Bessel function, spherical harmonic, and $\mathcal{R}$-tensor accumulation, along with \texttt{cuBLAS} for the final grid contraction. As shown in Table~\ref{tab:benchmarks}, this typically speeds up execution by a factor of $\sim 5$ to $6$, making the spherical method competitive with the FFT method in terms of speed, but with far fewer numerical challenges.

\section{Verification}\label{sec:Verification}

To verify that \texttt{SCarFFF} gives the correct results, we apply two key checks. The first, and simplest, is exclusive to the FFT method. As we are performing a Fourier transform, and have access to both the transition density and form factor, a natural check is that Parseval's theorem holds. That is, the transition density and form factor should satisfy the relation
\begin{equation}
    \int d^3r |\Phi^{g\to s}(\vec{r})|^2 = \int \frac{d^3q}{(2\pi)^3} |f_s(\vec{q})|^2.
\end{equation}
This check can be toggled with the \texttt{check\_parseval} flag when running in FFT mode. Ensuring that this equality holds can help to safeguard against numerical artifacts coming from grids with insufficient extent or resolution.

The second check is the oscillator strength reconstruction, which is available to all methods. This verification strategy checks that the oscillator strength computed from the tabulated form factor matches the \texttt{PySCF} output, which are computed reliably using TD-DFT. To compute the oscillator strength, we make use of the low momentum expansion of the form factor
\begin{equation}
    f_s(\vec{q}) \equiv \bra{\Psi_s} e^{i\vec{q}\cdot \vec{r}}\ket{\Psi_g} \simeq i\vec{q}\cdot \hvec{\mu} + \mathcal{O}(|\vec{q}|^2), 
\end{equation}
up to spin factors, with $\hvec{\mu} \equiv \langle \Psi_s | \vec r | \Psi_g \rangle$ the transition dipole moment, which in the basis of molecular orbitals $\varphi_p$ is given by
\begin{align}
\hvec \mu &= \sum_{pq} T_{pq}^{(s)} \langle \varphi_q | \vec r | \varphi_p \rangle.
\end{align}
This is related to the oscillator strength, $f_0$, by
\begin{equation}
    f_0 = \frac{2}{3}m_e \Delta E_{fi}|\hvec{\mu}|^2,
\end{equation}
where 
$\Delta E_{s}$ is the transition energy. 

For the real-valued transition density matrix, $\hvec\mu$ can be extracted from the Taylor expansion of $\text{Im}(f_s)$: 
\begin{align}
\text{Im}\,f_s(\vec q) &\simeq 0 + \vec q \cdot \nabla_q \text{Im}\,f_s(\vec q)\Big|_{ q \rightarrow 0} + \ldots 
\\
\hvec{\mu} &= \nabla_q \text{Im}\,f_s(\vec q)\Big|_{ q \rightarrow 0} ,
\end{align}
or equivalently from the quadratic moment of the squared form factor:
\begin{equation}
    f_0 = 2m_e \Delta E_{fi}\lim_{|\vec{q}|\to 0} \frac{\langle |f_s(\vec{q})|^2\rangle_\Omega}{|\vec{q}|^2}.
\end{equation}
In \texttt{SCarFFF} we perform a fit to the angular average of the squared form factor up to some user-defined $|\vec{q}|_\mathrm{max} \simeq 0.05 $ to $0.25\,\mathrm{keV}$,
\begin{equation}
    \langle |f_s(\vec{q})|^2\rangle_\Omega \simeq A + B |\vec{q}|^2 + C|\vec{q}|^4,
\end{equation}
and extract the coefficient $B = |\vec{\mu}|^2/3$. The agreement is naturally better for lower $|\vec{q}|_\mathrm{max}$, but is not always feasible due to coarse grids, particularly for the FFT. The first coefficient, $A$, is added to the fit to subtract any numerical artifacts of non-zero form factor at $\vec{q} = \vec{0}$, whilst $C$ accounts for the fact that we do not exactly take the $|\vec{q}| \to 0$ limit, but instead expand at finite, but small momenta. We find that for typical run parameters, \textit{e.g.} a threshold of $10^{-6}$, $\ell_\mathrm{max} = 24$, (if using the spherical method), and a $100^3$ $\vec{q}$ point grid, that our oscillator strengths match those of \texttt{PySCF} to the $10^{-4}$ level or better, provided $|\vec{q}|_\mathrm{max} = 0.05\,\mathrm{keV}$.

\begin{figure*}[t]
  \centering
  \begin{subfigure}[t]{0.33\textwidth}
  \centering
    \includegraphics[width=\linewidth]{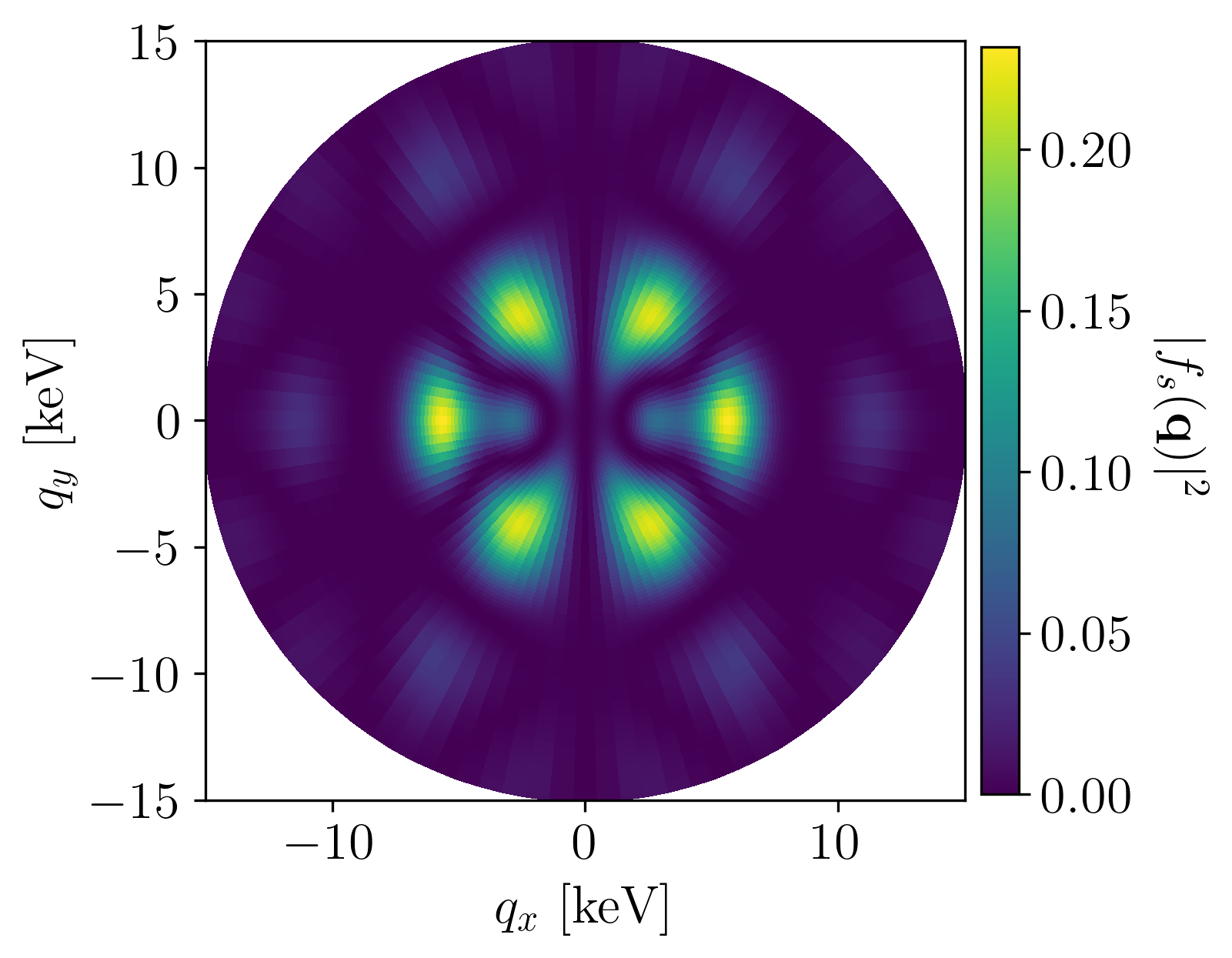}
    \caption{Spherical}
    \label{fig:first}
  \end{subfigure}\hfill
  \begin{subfigure}[t]{0.33\textwidth}
  \centering
    \includegraphics[width=\linewidth]{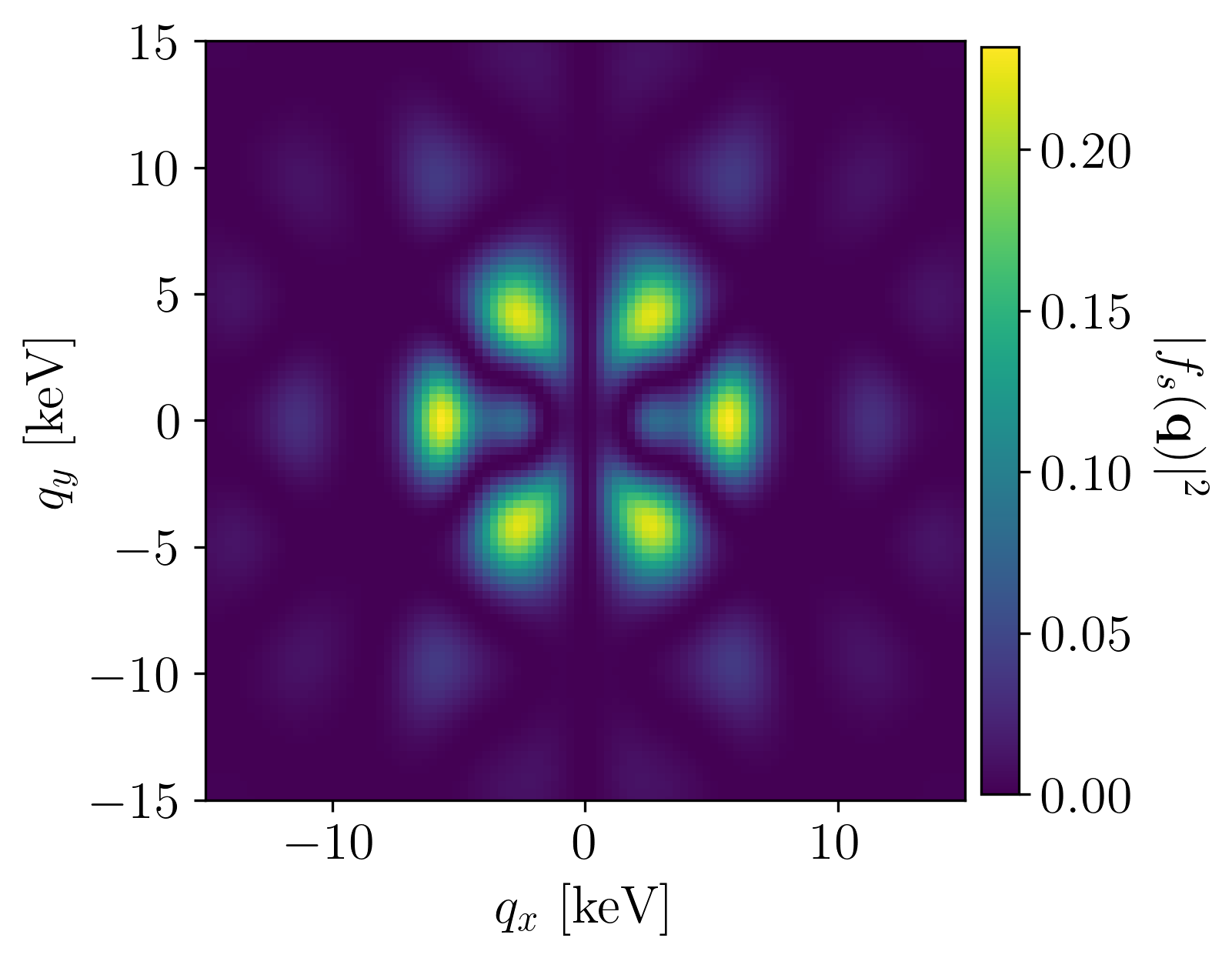}
    \caption{Cartesian}
    \label{fig:second}
  \end{subfigure}\hfill
  \begin{subfigure}[t]{0.33\textwidth}
  \centering
    \includegraphics[width=\linewidth]{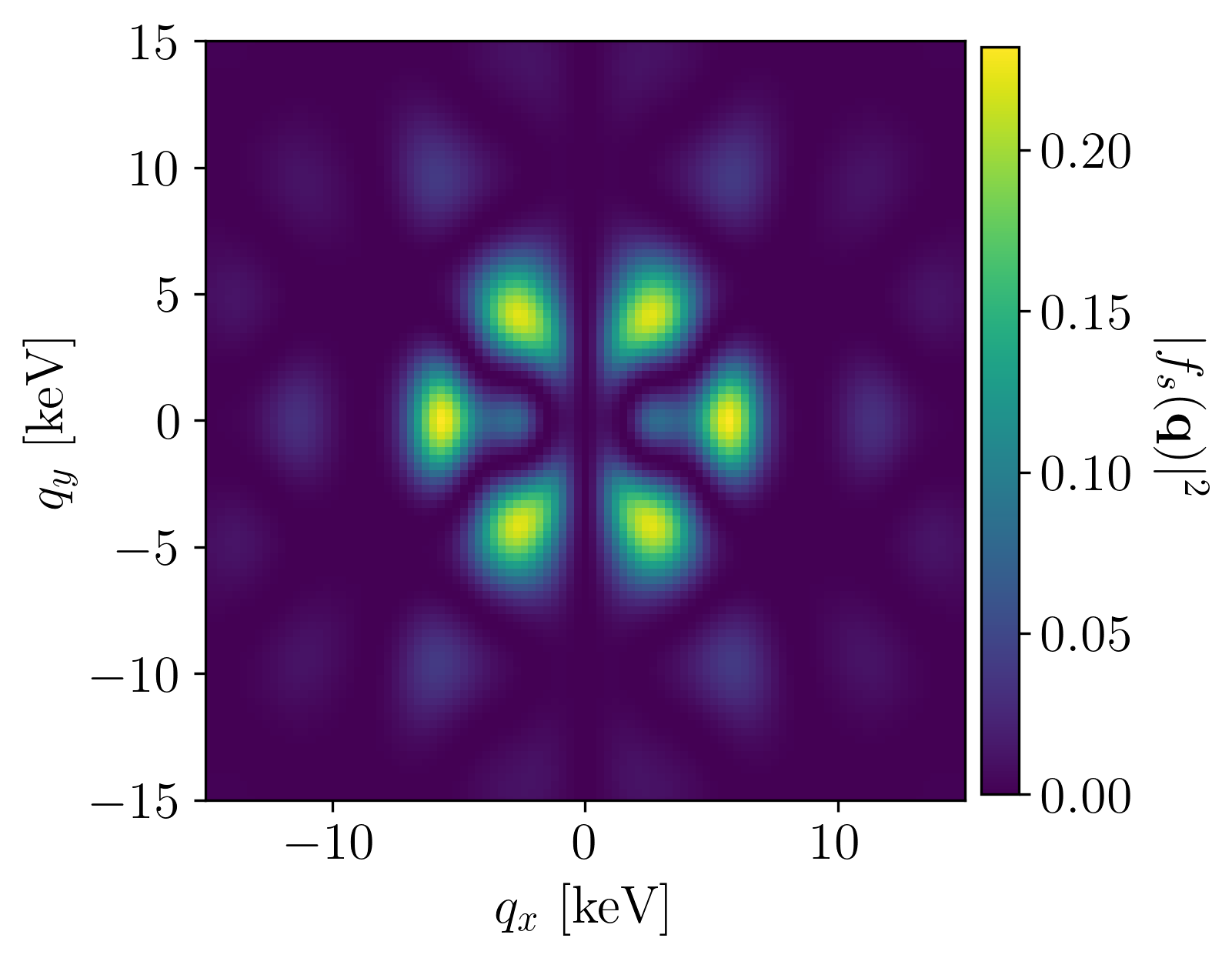}
    \caption{FFT}
    \label{fig:comparison}
  \end{subfigure}

  \caption{The squared form factor, $|f_s(\vec{q})|^2$, for the second excited state transition of p-xylene, using a) the spherical method with $\ell_\mathrm{max} = 24$, b) the Cartesian method, and c) the FFT method. We use the cc-pVDZ basis set, B3LYP exchange-correlation functional, and threshold $\epsilon = 10^{-6}$ for all three computations, and $N_{q_i} = 100$ grid points in each direction for both the spherical and Cartesian methods. To achieve the same accuracy with the FFT method, we have scaled up to $N_{q_i} = 400$. At $N_{q_i} = 100$ grid points, the FFT method disagrees by $\sim10\%$ with the analytic integration methods.}
  \label{fig:triple}
\end{figure*}

In Fig.~\ref{fig:comparison} we show the form factor for the second excited state p-xylene computed using \texttt{SCarFFF}, using each of the three methods. This demonstrates the self-consistency of our package, \textit{i.e.} that the results from all of our methods agree. We draw particular attention to the FFT method however, which had to be computed on a grid with $25 \,\mathrm{keV}$ extent, and a resolution of $0.125\,\mathrm{keV}$ to converge to the same result as the analytic methods, for a total of $4^3 = 64$ times more grid points.

\section{Conclusions}
\label{sec:Conclusions}

\texttt{SCarFFF} offers three complementary pathways for the speedy evaluation of molecular form factors. The Cartesian and spherical analytic methods can both be run without loss of precision, generating reliable results in a fraction of the time spent on the physical chemistry calculation.
When speed rather than precision is the priority, in particular for CPU-only runs, the FFT-based method provides the fastest analysis. 
Our numerical method \texttt{SCarFFF} is publicly available as a package in Julia, and can be found at
\begin{align*}
\texttt{https://github.com/jdshergold/SCarFFF}
\end{align*}

Although \texttt{SCarFFF} uses \texttt{PySCF} for its TD-DFT calculation, the analytic method we present in this paper can be used more generally, for any system where the molecular geometry and the GTO-basis transition density matrix are known.
This makes our methods highly adaptable, and compatible in principle with any other electronic structure computation.
The present version of \texttt{SCarFFF} captures the spin-independent molecular form factor for electronic transitions between spin singlet states, but we intend to broaden our scope to include both types of spin dependence in future work. 

\section*{Acknowledgments}
We would like to thank Juri Smirnov, Cameron Cook, Yonatan Kahn, Louis Hamaide, and Robert McGehee for helpful conversations. J.D.S thanks Clare Burrage and Jonas Spinner for helpful discussions and invaluable moral support towards the end of this project. The work of B.L.\ was supported in part by the U.S. Department of Energy under Grant Number DE-SC0011640. J.D.S. is supported through funding from the UK Research and Innovation Future Leader Fellowship MR/Y018656/1.

\appendix

\section{Atomic Orbital Definitions} \label{appx:basis}

Each of our atomic orbitals is written as a sum of primitive basis functions as
\begin{equation}
    \phi_{\alpha}(\vec{r} )= N_\alpha \sum_\mu \chi_\mu (\vec{r}),
\end{equation}
where $\alpha,\beta \dots$ are used to denote orbital specific quantities, $\mu,\nu\dots$ are to denote primitive specific quantities, and $N_\alpha$ is a normalisation factor. Each of these primitives is itself a sum of Cartesian terms
\begin{equation} \label{eq:defprimitive}
    \chi_\mu(\vec{r}) = d_\mu \xi_\mu \exp\left(-\frac{r^2}{2\sigma_\mu^2}\right) \sum_{i\in \mu} k_{\mu i} x^{a_{\mu i}} y^{b_{\mu i}} z^{c_{\mu i}},
\end{equation}
with $\sigma_{\mu}$ the Gaussian width of the primitive, and $d_\mu$ an overall scaling coefficient. Each choice of basis set (e.g.~6-31g* or cc-pVDZ) is defined by its values of $d_\mu$ and $\sigma_\mu$.  The coefficients $k_i$ are the \textit{Cartesian} prefactors, which could be \textit{e.g.} $-1$ for the $y^2$ term in the 3$d_{x^2-y^2}$ orbital. Similarly, $a_i, b_i,$ and $c_i$ denote the degree of $x,y,$ and $z$ for orbital, respectively. We will denote objects the Cartesian term level with the indices $i$ and $j$ throughout.

The normalisation coefficients are fixed such that individual primitives are normalised to
\begin{equation}
    \int d^3r \, |\chi_{\mu}(\vec{r})|^2 = d_\mu^2,
\end{equation}
whilst contracted orbitals are normalised to unity. This amounts to
\begin{widetext}
    \begin{align} \label{eq:GTOnormalisation}
    \xi_\mu &= \left[\sum_{i,j\in \mu} k_{\mu i} k_{\mu j} \prod_{K \in \{\alpha,\beta,\gamma\}} \sigma_\mu^{1 + K_{ij}^{\mu\mu}} \Gamma\left(\frac{K_{ij}^{\mu\mu} + 1}{2}\right)\right]^{-\frac{1}{2}},\\
    N_\alpha &= \left[\sum_{\mu,\nu} d_\mu d_\nu \xi_\mu \xi_\nu \sum_{i\in \mu,j\in \nu} k_{\mu i} k_{\nu j} \prod_{K \in \{\alpha,\beta,\gamma\}} (\sqrt{2}\,\sigma_{\mu\nu})^{1 + K_{ij}^{\mu\nu}} \Gamma\left(\frac{K_{ij}^{\mu\nu} + 1}{2}\right)\right]^{-\frac{1}{2}},
\end{align}
\end{widetext}
with $\alpha_{ij}^{\mu\nu} = a_{\mu i} + a_{\nu j}$, $\beta_{ij}^{\mu\nu} = b_{\mu i} + b_{\nu j}$ and $\gamma_{ij}^{\mu\nu} = c_{\mu i} + c_{\nu j}$, and $\sigma_{\mu\nu}$ as defined in~\eqref{eq:sigmaTidbits}.

As a concrete example, we take a $p_z$ orbital, which has just a single Cartesian term with $(a, b, c, k) = (0,0,1,1)$, \textit{i.e.}
\begin{equation}
    \chi_\mu(\vec{r}) = d_\mu \xi_\mu \, z \exp\left(-\frac{r^2}{2\sigma_\mu^2}\right), 
\end{equation}
such that
\begin{equation}
    \xi_\mu = \left(\frac{1}{2} \pi^\frac{3}{2} \sigma_\mu^5\right)^{-\frac{1}{2}}.
\end{equation}
In a toy orbital with two primitives, labeled with $\mu \in \{0,1\}$, this would then have an overall normalisation
\begin{equation}
    N_\alpha = \left[d_0^2 + d_1^2 + 2d_0 d_1 \left(\frac{2\sigma_0 \sigma_1}{\sigma_0^2 +\sigma_1^2}\right)^{\frac{5}{2}}\right]^{-\frac{1}{2}}.
\end{equation}

\section{Monomial-to-spherical Coefficients}\label{sec:thetaCoefficients}
The key feature that sets the spherical grid method above the Cartesian grid method is the ability to decouple the $|\vec{q}|$ grid computation from the angular, $(q_\theta,q_\phi)$, grid computation. However, in order to go from a Cartesian to spherical grid, we need to map monomials in $q_x, q_y,$ and $q_z$ onto a basis of spherical harmonics. That is, we require the coefficients $\Theta_{\lambda\mu}^{uvw}$ satisfying
\begin{equation}
    q_x^u q_y^v q_z^z = |\vec{q}|^n\sum_{\lambda,\mu} \Theta_{\lambda\mu}^{uvw} Y_{\lambda}^\mu(\hat{q}),
\end{equation}
where $n = u + v + w$. The first step towards this goal is to write the Cartesian momentum components in terms of the $\ell = 1$ spherical harmonics as
\begin{equation}\label{eq:qiToYlm}
    q_i = |\vec{q}|\sqrt{\frac{4\pi}{3}} \sum_{m} c_{im} Y_{1}^m(\hat q),
\end{equation}
where the coefficients follow from the transformation from Cartesian coordinates to the spherical basis
\begin{align}
    c_{xm} &= \frac{1}{\sqrt{2}}\{1,0,-1\}, \\
    c_{ym} &= \frac{i}{\sqrt{2}}\{1,0,1\}, \\
    c_{zm} &= \{0,1,0\},
\end{align}
with the entries corresponding to $m = \{-1,0,1\}$. This decomposition into spherical tensor operators reveals several properties of the $\Theta$ coefficients. First, as only $q_x$ and $q_y$ can raise or lower angular momenta, $\mu$ must have the same parity as $u + v$, and must also satisfy $|\mu| \leq u + v$. Second, as the coupling of $n$ rank-$1$ spherical tensor operators can result in at most a rank-$n$ tensor, we must have $\lambda \leq n$. Finally, and more subtly, we know that under spatial inversions, $\prod_{i=1}^{n} q_i \to (-1)^n \prod_{i=1}^{n} q_i$, whilst $Y^{\mu}_{\lambda}(\hat{q}) \to (-1)^{\lambda} Y^{\mu}_{\lambda}(\hat{q})$. As a result, only coefficients where $\lambda$ and $n$ have the same parity survive. The full set of selection rules is then
\begin{align}
     0 \leq \lambda \leq n, &\qquad \lambda \equiv n \;(\mathrm{mod}\  2), \\
    |\mu| \leq \mathrm{min}\{\lambda, u+v\}, &\qquad \mu\equiv u+v \;(\mathrm{mod}\  2),
\end{align}
which can be leveraged to significantly speed up the computation of the $\Theta$ coefficients. 

Using~\eqref{eq:qiToYlm}, we can rewrite the monomial in $q_i$ as
\begin{widetext}
\begin{equation}
    q_x^u q_y^v q_z^w = |\vec{q}|^n \left(\frac{4\pi}{3}\right)^\frac{n}{2} \sum_{\{m_i\}} \left[\left(\prod_{i=1}^u c_{xm_i}\right)\left(\prod_{i=u+1}^{u+v} c_{ym_i}\right)\left(\prod_{i=u+v+1}^{n} c_{zm_i}\right)\left(\prod_{i=1}^{n} Y_{1}^{m_i}(\hat{q})\right)\right],
\end{equation}
\end{widetext}
where the sum runs over all combinations of $n$ magnetic quantum numbers, $m_i \in \{-1,0,1\}$. The next step is to couple the $\ell = 1$ spherical harmonics to yield an expression of the form
\begin{equation}\label{eq:Kcoefficients}
    \prod_{i=1}^{n} Y_{1}^{m_i}(\hat{q}) = \sum_{\lambda} \mathcal{K}_{\lambda}^n(\{m_i\}) Y_{\lambda}^{\mu_n}(\hat q),
\end{equation}
with $\mu_n = \sum_{i=1}^{n} m_i$, and $\lambda  \in \{n, n-2, \dots\}$, which follows from our earlier parity arguments. The form of the $\mathcal{K}$ coefficients is found by repeatedly applying the spherical harmonic contraction rule given in~\eqref{eq:YYY}. This leads to the recursive definition of the $\mathcal{K}$ coefficients
\begin{align}
    \mathcal{K}_{\lambda}^1(\{m_i\}) &= \delta_{\lambda 1}, \\
    \mathcal{K}_{\lambda}^n(\{m_i\}) &= \sum_L \mathcal{G}_{L\mu_{n-1},1m_n}^{\lambda \mu_n} \mathcal{K}_{L}^{n-1}(\{m_i\}),
\end{align}
where in the second line, $L \in \{n-1, n-3, \dots\}$, and $m_n = \mu_n - \mu_{n-1}$. Applying \eqref{eq:Kcoefficients} and reordering the sums yields the desired form of the $\Theta$ coefficients
\begin{widetext}
\begin{equation}
    \Theta_{\lambda\mu}^{uvw} = \left(\frac{4\pi}{3}\right)^\frac{n}{2} \sum_{\{m_i\}} \left[\left(\prod_{i=1}^u c_{xm_i}\right)\left(\prod_{i=u+1}^{u+v} c_{ym_i}\right)\left(\prod_{i=u+v+1}^{n} c_{zm_i}\right) \mathcal{K}_{\lambda}^n(\{m_i\}) \delta_{\mu_n \mu}\right].
\end{equation}
\end{widetext}

\bibliography{references}

@article{Blanco:2021hlm,
    author = "Blanco, Carlos and Kahn, Yonatan and Lillard, Benjamin and McDermott, Samuel D.",
    title = "{Dark Matter Daily Modulation With Anisotropic Organic Crystals}",
    eprint = "2103.08601",
    archivePrefix = "arXiv",
    primaryClass = "hep-ph",
    reportNumber = "FERMILAB-PUB-21-066-T",
    doi = "10.1103/PhysRevD.104.036011",
    journal = "Phys. Rev. D",
    volume = "104",
    pages = "036011",
    year = "2021"
}

@article{Giffin:2025hdx,
    author = "Giffin, Pierce and Lillard, Benjamin and Munbodh, Pankaj and Yu, Tien-Tien",
    title = "{Simplified Spin Dependence in Dark Matter Direct Detection}",
    eprint = "2511.10764",
    archivePrefix = "arXiv",
    primaryClass = "hep-ph",
    month = "11",
    year = "2025"
}

@article{Lillard:2023qlx,
    author = "Lillard, Benjamin",
    title = "{Partial Rate Matrix for Dark Matter Scattering}",
    eprint = "2310.01480",
    archivePrefix = "arXiv",
    primaryClass = "hep-ph",
    doi = "10.1103/PhysRevLett.134.221003",
    journal = "Phys. Rev. Lett.",
    volume = "134",
    number = "22",
    pages = "221003",
    year = "2025"
}

@article{Lillard:2023cyy,
    author = "Lillard, Benjamin",
    title = "{Wavelet-harmonic integration methods}",
    eprint = "2310.01483",
    archivePrefix = "arXiv",
    primaryClass = "hep-ph",
    doi = "10.1103/PhysRevD.111.123006",
    journal = "Phys. Rev. D",
    volume = "111",
    number = "12",
    pages = "123006",
    year = "2025"
}

@article{Lillard:2025aim,
    author = "Lillard, Benjamin and Radick, Aria",
    title = "{Vector Spaces for Dark Matter (VSDM): Fast Direct Detection Calculations with Python and Julia}",
    eprint = "2502.17547",
    archivePrefix = "arXiv",
    primaryClass = "hep-ph",
    month = "2",
    year = "2025"
}

@incollection{casida1995time,
  title={Time-dependent density functional response theory for molecules},
  author={Casida, Mark E},
  booktitle={Recent Advances In Density Functional Methods: (Part I)},
  pages={155--192},
  year={1995},
  publisher={World Scientific}
}

@article{Hochberg:2021pkt,
    author = "Hochberg, Yonit and Kahn, Yonatan and Kurinsky, Noah and Lehmann, Benjamin V. and Yu, To Chin and Berggren, Karl K.",
    title = "{Determining Dark-Matter{\textendash}Electron Scattering Rates from the Dielectric Function}",
    eprint = "2101.08263",
    archivePrefix = "arXiv",
    primaryClass = "hep-ph",
    reportNumber = "FERMILAB-PUB-21-128-AE",
    doi = "10.1103/PhysRevLett.127.151802",
    journal = "Phys. Rev. Lett.",
    volume = "127",
    number = "15",
    pages = "151802",
    year = "2021"
}

@article{Knapen:2021run,
    author = "Knapen, Simon and Kozaczuk, Jonathan and Lin, Tongyan",
    title = "{Dark matter-electron scattering in dielectrics}",
    eprint = "2101.08275",
    archivePrefix = "arXiv",
    primaryClass = "hep-ph",
    reportNumber = "CERN-TH-2021-013",
    doi = "10.1103/PhysRevD.104.015031",
    journal = "Phys. Rev. D",
    volume = "104",
    number = "1",
    pages = "015031",
    year = "2021"
}

@article{Lasenby:2021wsc,
    author = "Lasenby, Robert and Prabhu, Anirudh",
    title = "{Dark matter{\textendash}electron scattering in materials: Sum rules and heterostructures}",
    eprint = "2110.01587",
    archivePrefix = "arXiv",
    primaryClass = "hep-ph",
    doi = "10.1103/PhysRevD.105.095009",
    journal = "Phys. Rev. D",
    volume = "105",
    number = "9",
    pages = "095009",
    year = "2022"
}

@article{Boyd:2022tcn,
    author = "Boyd, Christian and Hochberg, Yonit and Kahn, Yonatan and Kramer, Eric David and Kurinsky, Noah and Lehmann, Benjamin V. and Yu, To Chin",
    title = "{Directional detection of dark matter with anisotropic response functions}",
    eprint = "2212.04505",
    archivePrefix = "arXiv",
    primaryClass = "hep-ph",
    doi = "10.1103/PhysRevD.108.015015",
    journal = "Phys. Rev. D",
    volume = "108",
    number = "1",
    pages = "015015",
    year = "2023"
}

@article{Sun2020PySCF,
  author  = {Sun, Qiming and Zhang, Xing and Banerjee, Samragni and Bao, Peng and Barbry, Marc and Blunt, Nick S. and Bogdanov, Nikolay A. and Booth, George H. and Chen, Jia and Cui, Zhi-Hao and Eriksen, Janus Juul and Gao, Yang and Guo, Sheng and Hermann, Jan and Hermes, Matthew R. and Koh, Kevin and Koval, Peter and Lehtola, Susi and Li, Zhendong and Liu, Junzi and Mardirossian, Narbe and McClain, James D. and Motta, Mario and Mussard, Bastien and Pham, Hung Q. and Pulkin, Artem and Purwanto, Wirawan and Robinson, Paul J. and Ronca, Enrico and Sayfutyarova, Elvira and Scheurer, Maximilian and Schurkus, Henry F. and Smith, James E. T. and Sun, Chong and Sun, Shi-Ning and Upadhyay, Shiv and Wagner, Lucas K. and Wang, Xiao and White, Alec and Whitfield, James Daniel and Williamson, Mark J. and Wouters, Sebastian and Yang, Jun and Yu, Jason M. and Zhu, Tianyu and Berkelbach, Timothy C. and Sharma, Sandeep and Sokolov, Alexander and Chan, Garnet Kin-Lic},
  title   = {Recent developments in the {PySCF} program package},
  journal = {J. Chem. Phys.},
  year    = {2020},
  volume  = {153},
  number  = {2},
  pages   = {024109},
  doi     = {10.1063/5.0006074}
}

@article{Wang2016geomeTRIC,
  author  = {Wang, Lee-Ping and Song, Chenchen},
  title   = {Geometry optimization made simple with translation and rotation coordinates},
  journal = {J. Chem. Phys.},
  year    = {2016},
  volume  = {144},
  number  = {21},
  pages   = {214108},
  doi     = {10.1063/1.4952956}
}

@misc{rdkit_overview,
  author       = {Landrum, Greg and others},
  title        = {The {RDKit} Documentation: {An} overview of the {RDKit}},
  howpublished = {\url{https://www.rdkit.org/docs/Overview.html}},
  note         = {Accessed: 2025-12-17},
  year         = {2025}
}

@article{FrigoJohnson2005FFTW3,
  author  = {Frigo, Matteo and Johnson, Steven G.},
  title   = {The Design and Implementation of {FFTW3}},
  journal = {Proceedings of the IEEE},
  year    = {2005},
  volume  = {93},
  number  = {2},
  pages   = {216--231},
  doi     = {10.1109/JPROC.2004.840301}
}

@article{nto_martin,
    author = {Martin, Richard L.},
    title = {Natural transition orbitals},
    journal = {The Journal of Chemical Physics},
    volume = {118},
    number = {11},
    pages = {4775-4777},
    year = {2003},
    month = {03},
    issn = {0021-9606},
    doi = {10.1063/1.1558471},
    url = {https://doi.org/10.1063/1.1558471},
    eprint = {https://pubs.aip.org/aip/jcp/article-pdf/118/11/4775/19210043/4775\_1\_online.pdf},
}

@article{Zwicky:1933gu,
    author = "Zwicky, F.",
    title = "{Die Rotverschiebung von extragalaktischen Nebeln}",
    doi = "10.1007/s10714-008-0707-4",
    journal = "Helv. Phys. Acta",
    volume = "6",
    pages = "110--127",
    year = "1933"
}

@article{Rubin:1970zza,
    author = "Rubin, Vera C. and Ford, Jr., W. Kent",
    title = "{Rotation of the Andromeda Nebula from a Spectroscopic Survey of Emission Regions}",
    doi = "10.1086/150317",
    journal = "Astrophys. J.",
    volume = "159",
    pages = "379--403",
    year = "1970"
}

@article{Essig:2015cda,
    author = "Essig, Rouven and Fernandez-Serra, Marivi and Mardon, Jeremy and Soto, Adrian and Volansky, Tomer and Yu, Tien-Tien",
    title = "{Direct Detection of sub-GeV Dark Matter with Semiconductor Targets}",
    eprint = "1509.01598",
    archivePrefix = "arXiv",
    primaryClass = "hep-ph",
    doi = "10.1007/JHEP05(2016)046",
    journal = "JHEP",
    volume = "05",
    pages = "046",
    year = "2016"
}

@article{Trickle:2019nya,
    author = "Trickle, Tanner and Zhang, Zhengkang and Zurek, Kathryn M. and Inzani, Katherine and Griffin, Sin{\'e}ad M.",
    title = "{Multi-Channel Direct Detection of Light Dark Matter: Theoretical Framework}",
    eprint = "1910.08092",
    archivePrefix = "arXiv",
    primaryClass = "hep-ph",
    doi = "10.1007/JHEP03(2020)036",
    journal = "JHEP",
    volume = "03",
    pages = "036",
    year = "2020"
}

@article{Hochberg:2025rjs,
    author = "Hochberg, Yonit and Khalaf, Majed and Lenoci, Alessandro and Ovadia, Rotem",
    title = "{Determining (All) Dark Matter-Electron Scattering Rates From Material Properties}",
    eprint = "2510.25835",
    archivePrefix = "arXiv",
    primaryClass = "hep-ph",
    month = "10",
    year = "2025"
}

@article{Catena:2024rym,
    author = "Catena, Riccardo and Spaldin, Nicola A.",
    title = "{Linear response theory for light dark matter-electron scattering in materials}",
    eprint = "2402.06817",
    archivePrefix = "arXiv",
    primaryClass = "hep-ph",
    doi = "10.1103/PhysRevResearch.6.033230",
    journal = "Phys. Rev. Res.",
    volume = "6",
    number = "3",
    pages = "033230",
    year = "2024"
}

@article{Berlin:2025uka,
    author = "Berlin, Asher and Millar, Alexander J. and Trickle, Tanner and Zhou, Kevin",
    title = "{Determining spin-dependent light dark matter rates from neutron scattering}",
    eprint = "2504.02927",
    archivePrefix = "arXiv",
    primaryClass = "hep-ph",
    reportNumber = "FERMILAB-PUB-25-0221-T",
    doi = "10.1103/4yxf-lkr6",
    journal = "Phys. Rev. D",
    volume = "112",
    number = "3",
    pages = "035021",
    year = "2025"
}

\end{document}